\documentclass[preprint,12pt]{elsarticle}

\usepackage[utf8]{inputenc}
\usepackage{xcolor}

\usepackage{lineno}
\usepackage[letterpaper,top=2cm,bottom=2cm,left=2cm,right=2cm,marginparwidth=1.75cm]{geometry}

\usepackage{amsmath}
\usepackage{palatino,graphicx,wrapfig}
\usepackage[small,it]{caption}
\usepackage{amsmath}
\usepackage{amssymb,bm}
\usepackage{parskip}

\journal{}

\begin{document}

% \title{Timescale-eliminating neural ODE surrogates for advection-dominated dynamical systems}
\title{Understanding Latent Timescales in Neural Ordinary Differential Equation Models for Advection-Dominated Dynamical Systems}
\author[label1,label2]{Ashish S. Nair}
\author[label1]{Shivam Barwey\corref{cor1}}
\author[label1]{Pinaki Pal}
\author[label2]{Jonathan F. MacArt}
\author[label1]{Troy Arcomano}
\author[label1,label3]{Romit Maulik}

\cortext[cor1]{Corresponding author. E-mail address: {sbarwey@anl.gov} (S. Barwey).}

\affiliation[label1]{organization={Argonne National Laboratory},
            addressline={9700 South Cass Avenue},
            city={Lemont},
            postcode={60439},
            state={IL},
            country={USA}}

\affiliation[label2]{organization={University of Notre Dame},
            addressline={Holy Cross Dr},
            city={Note Dame},
            postcode={46556},
            state={IN},
            country={USA}}
            
\affiliation[label3]{organization={Pennsylvania State University},
            addressline={E327 Westgate Building},
            city={University Park},
            postcode={16802},
            state={PA},
            country={USA}}

\begin{abstract}
The neural ordinary differential equation (ODE) framework has shown considerable promise in recent years in developing highly accelerated surrogate models for complex physical systems characterized by partial differential equations (PDEs). For PDE-based systems, state-of-the-art neural ODE strategies leverage a two-step procedure to achieve this acceleration: a nonlinear dimensionality reduction step provided by an autoencoder, and a time integration step provided by a neural-network based model for the resultant latent space dynamics (the neural ODE). This work explores the applicability of such autoencoder-based neural ODE strategies for PDEs in which advection terms play a critical role. More specifically, alongside predictive demonstrations, physical insight into the sources of model acceleration (i.e., how the neural ODE achieves its acceleration) is the scope of the current study. Such investigations are performed by quantifying the effects of both autoencoder and neural ODE components on latent system time-scales using eigenvalue analysis of dynamical system Jacobians. To this end, the sensitivity of various critical training parameters -- de-coupled versus end-to-end training, latent space dimensionality, and the role of training trajectory length, for example -- to both model accuracy and the discovered latent system timescales is quantified. This work specifically uncovers the key role played by the training trajectory length (the number of rollout steps in the loss function during training) on the latent system timescales: larger trajectory lengths correlate with an increase in limiting neural ODE time-scales, and optimal neural ODEs are found to recover the largest time-scales of the full-order (ground-truth) system. Demonstrations are performed across fundamentally different unsteady fluid dynamics configurations influenced by advection: (1) the Kuramoto-Sivashinsky equations (2) Hydrogen-Air channel detonations (the compressible reacting Navier-Stokes equations with detailed chemistry), and (3) 2D Atmospheric flow.

\end{abstract}

\maketitle

\section{Introduction} \label{Sec: Intro}

Numerical solutions of partial differential equations (PDEs), if available, can be used by domain scientists to not only probe complex physical behavior at unprecedented levels of accuracy and detail, but also to accelerate design optimization workflows for engineering devices. Simulating fluid dynamics, for example, requires numerically solving the Navier-Stokes PDEs \cite{moin_dns}. At industrial operating conditions of interest -- those that characterize flows over aircraft wings, in scramjets, and in gas turbine combustion chambers, for instance -- these equations admit multi-scale and multi-physics behavior stemming from interactions between turbulence, shock waves, and potential chemical reactions. As a result, to generate reliable simulations of such devices, PDE solution procedures need to resolve all length-scales and time-scales that characterize these physical phenomena. These spatiotemporal resolution requirements render (a) long-time direct numerical simulation (DNS) of the aforementioned systems infeasible, and (b) real-time simulation-based actuation or control strategies impractical, despite recent advances in supercomputing technology and physics simulation hardware \cite{venkat_emergingtrends,exascale}. 

Reduced-order models (ROMs) are a class of modeling approaches that seek to eliminate intrinsic costs associated with physics-based simulations \cite{lucia2004reduced}, with the goal of enabling long-time simulation capability. The general ROM objective is to achieve drastic reduction in the PDE-derived dynamical system, which is the high-dimensional nonlinear set of ordinary differential equations (ODEs) produced from PDE discretization. In the context of fluid flows, this dynamical system describes the evolution of a so-called \textit{state vector} described by turbulent fluid density, momentum, and species concentration fields on a grid, and can readily reach on the order of hundreds of millions (and even billions) of degrees-of-freedom. As such, to achieve model-based reduction, ROMs typically leverage a two-step approach: an offline projection step which generates the reduced representation of the state vector, and an online forecasting step which generates the time-evolution of the reduced state. Within this scope, both physics-based and data-based ROMs can be constructed. 

Physics-based ROMs for fluid flows typically leverage linear projection operations to resolve only the large scales, which are assumed to contain a majority of the system energy; the effect of the unresolved (small) scales on the resolved dynamics is then modeled. A classic example is large-eddy simulation (LES) \cite{fox_les,fureby_les}, where the projection operation (the mechanism for dimensionality reduction) is a non-invertible spatial filter \cite{ideal_les}, and the forecasting step requires solving a filtered version of the Navier-Stokes equations. The LES closure model -- which captures the effects of the unresolved (small) scales on the resolved dynamics -- can come from either phenomenological algebraic relations \cite{dynamic_model} or statistical closures \cite{ideal_les,adrian_1988}. Other physics-based ROMs achieve dimensionality reduction in different ways, such as through inertial manifold assumptions \cite{foias_aim,maryam_aim}, Koopman operator theory \cite{parish_2017}, and more intrusive alterations to the governing PDEs (e.g., flamelet models used in combustion modeling \cite{pitsch_arfm} and two-dimensional turbulence used in climate modeling \cite{2d_turb}). 

Data-based ROMs, on the other hand, rely on samples of the full-dimensional state vector (i.e., fluid flow snapshots) to produce projection operators in an optimization (or training) step. Although this incurs a large offline computational cost not present in physics-based counterparts, these methods have been shown in recent years to produce significantly larger levels of dimensionality reduction, such that this training cost can be offset by immense levels of speedup achieved during the forecasting stage \cite{kutz2017deep,karthik_arfm,venkat_emergingtrends}. In the context of data-based ROMs, the reduced space is termed the \textit{latent space}. The initial projection step produces the (reduced) latent variables, and the forecasting step requires modeling the dynamics of these latent variables. 

Data-based ROMs have a history spanning several decades, and their ability to capture complex physics contained in training datasets has led to their increased adoption. Methods rooted in modal decomposition -- including proper orthogonal decomposition (POD) \cite{berkooz_pod}, dynamic mode decomposition (DMD) \cite{schmid_dmd_arfm}, resolvent analysis \cite{towne2018spectral}, and cluster-based methods \cite{gunzburger_cvt}, among others -- derive basis functions directly from data, which translates to latent spaces generated by linear projection operations. The properties of the resultant latent variables naturally vary based on the method used to produce the basis. In POD, a basis is produced from eigenvectors of the covariance matrix of the data, resulting in latent variables that are disentangled and optimally preserve the system energy \cite{berkooz_pod}. In DMD and resolvent analysis, the linear projection generates latent variables described by characteristic frequencies, similar to traditional Fourier-based methods \cite{schmid_dmd_arfm}. Cluster-based methods use data partitioning strategies to produce latent variables that are symbolic encodings of the system state \cite{kaiser_crom,shivam_timeaxis}. The evolution of latent variables in all of these methods can be modeled in a physics-derived manner through Galerkin projection of the basis onto the underlying PDEs \cite{pod_galerkin}, or in a data-based manner through the utilization of machine learning methods \cite{romit_spod}. 

All of the methods discussed above leverage linear projection operations to achieve reduction. These methods have been successfully used to produce ROMs for diffusion-dominated problems, such as turbulent flows in canonical configurations and simpler PDEs without advection terms, but face difficulites for advection-dominated problems (e.g., high Reynolds number turbulent flows and shock-containing flows) characterized by slow decaying Kolmogorov n-width~\cite{ohlberger2015reduced, LEE2020108973}. 

% Non-linear: advection dominated 
The utilization of data-based ROMs relying on nonlinear projection operations have been shown to overcome these limitations, extending both compression and forecasting capabilities of linear approaches \cite{romit_cnn_advection,lee2020model,glaws2020deep}. The backbone of these methods is the autoencoder, a compression approach that leverages the expressive power of neural networks to generate robust latent space representations \cite{autoencoder_review}. Autoencoders rely on two components: the encoder, which serves as the nonlinear projection that moves the high-dimensional state into its latent representation, and the decoder, which seeks to undo the action of the encoder by recovering the full state from the latent variables. Due to the generalizable nature of neural networks, autoencoders can take many forms tailored to the application at hand: for example, architectures can leverage multi-layer perceptrons \cite{bethany_koopman}, convolutional neural networks \cite{romit_cnn_advection,lee2020model}, graph neural networks \cite{gunzburger_gnn,shivam_gnn}, and transformers \cite{liu2019transformer}. 

The success of neural network based autoencoders has driven efforts to create purely data-based predictive models in latent space that offer unprecedented levels of speedup over physics-based counterparts \cite{lee_parish_2020}. A core advantage is that such models can leverage real-world observations (e.g., from experimental diagnostics and operational sensor streams), which is critical for applications that are too expensive to simulate directly or do not have a solidified set of governing equations ~\cite{xu2020multi, romit_cnn_advection, yu2019non, fresca2021comprehensive, duraisamy2015new, hesthaven2018non}. The goal of these forecasting models is to operate in concert with the autoencoder by using data to learn the dynamics of the latent variables. The latent dynamical systems can be modeled using nonlinear \cite{romit_lstm,reddy_rnn} or linear \cite{bethany_koopman,kaiser_crom} surrogate forecasting models; such models have been used to accelerate advection dominated fluid flow \cite{romit_cnn_advection}, chaotic dynamical systems \cite{bethany_koopman}, and stiff chemical kinetic simulations \cite{ji2021stiff}. These methods, however -- e.g., those based on strategies like recurrent neural networks (RNNs) \cite{romit_cnn_advection}, residual networks (ResNets) \cite{jiang2021deep}, and latent Koopman representations \cite{yeung2019learning} -- typically learn the latent dynamics in the context of discrete and explicit temporal discretization, which can limit predictive capability. 

The goal of this study is to develop surrogate models for advection-dominated systems in the neural ODE framework \cite{chen2018neural}, which offers unique advantages over the above mentioned approaches. More specifically, the objective here is twofold: (1) to develop a neural ODE based latent dynamics model for advection-dominated problems in a purely data-based setting, and (2) to conduct a detailed analysis of the discovered dynamics in latent space using timescale characterizations. Before outlining the specific contributions of this work, the neural ODE strategy as it relates to these objectives is first summarized.

Since its introduction in Ref.~\cite{chen2018neural}, the neural ODE strategy has been cemented as a powerful scientific machine learning tool to model the evolution of dynamical systems using neural networks. Instead of directly enforcing discrete temporal representations (e.g., as used in residual networks \cite{jiang2021deep} or recurrent neural networks \cite{reddy_rnn}), neural ODEs learn a continuous representation of nonlinear system dynamics. In other words, the \textit{instantaneous} right-hand-side is modeled as a nonlinear function via a neural network, allowing the framework to leverage existing, vetted time-integration schemes to execute the forecasting step (the time integration scheme is separated from the dynamics model) \cite{chemnode,Finlay2020,inproceedings, kumar2023posteriori, böttcher2024controlmedicaldigitaltwins}. {A prior version of neural ODEs, integrating feedforward neural networks with numerical methods for dynamical system identification, used a Runge–Kutta neural network framework \cite{Wang1998RungeKuttaNN}.} With autoencoder-provided latent spaces, the neural ODE conveniently outputs a functional form for the instantaneous rate-of-change of the latent variables, which is useful for reduced-order modeling applications. This autoencoder-based neural ODE strategy been used to develop accelerated surrogate models in a variety of physical applications, including chaotic fluid flows \cite{Dutta2021}, advection-dominated PDEs \cite{Linot2023,Wan2023}, and stiff chemical kinetics \cite{ Vijay23, kumar2024phychemnode}. {Moreover, neural ODE surrogates can be effectively combined with neural controllers to tackle complex control problems, as demonstrated in recent works on medical digital twins and optimal control of dynamical systems \cite{Mowlavi2021OptimalCO, Bttcher2021AIPO, Bottcher2022AIPontryagin}.} 

% What is missing in previous work? 
Ultimately, previous work has shown how the combination of autoencoders with neural ODEs can be used to generate highly accelerated reduced-order models of physical systems. Despite this, the source of acceleration in the overall modeling strategy remains unclear: the goal of achieving model accuracy from both forecasting and autoencoding perspectives often overshadows the need to identify the contribution of each of these components to the empirically observed model acceleration. Insights into the sources of acceleration provided by the neural ODE based ROM can lead to valuable physical and model-oriented insights, and is the scope of the current study. Recent work has observed the effect of smoothed latent trajectories for advection-dominated systems produced by neural ODE simulations, pointing to a relationship between model acceleration and instrinsic timescale elimination \cite{Wan2023}. Similar trends have been shown for neural ODE based surrogate models for stiff chemical kinetics \cite{Vijay23}. As such, a more rigorous and quantitative analysis of timescale elimination produced by both autoencoder and neural ODE components is warranted. Additionally, the effect of critical neural ODE training parameters -- such as the overall integration time used to evaluate model errors -- on the accuracy and degree of timescale elimination produced in the latent space has been largely unexplored in the literature. Lastly, although previous work has demonstrated application of neural ODE based surrogate models for simplified advection-dominated PDEs (e.g., Burger's equation), extension of this strategy to more complex shock-containing flows remains sparsely explored. To this end, the main contributions of this work are as follows:

\begin{itemize}

  \item Using eigenvalue analysis to quantify the fastest and slowest time-scales in the latent space.
  \item Evaluate the effects of training methodologies, network architecture hyperparameters, and training trajectory length ($n_t$) on accuracy and time-scale reduction in the latent space, highlighting that $n_t$ is a critical controlling parameter to determine the extent of time-scale reduction in the latent space.
  \item Extend the proposed framework to a challenging and highly advection-dominated test case, specifically 1D detonation wave propagation (considering stiff chemical kinetics), {and to a real-world 2D atmospheric dataset}, to validate and affirm the observed trends.
\end{itemize}

{ It is emphasized that while the approach of combining autoencoders for spatial compression with neural ODEs for learning latent dynamics has been previously explored, this work leverages the general autoencoder-based neural ODE strategy to facilitate a latent time-scale analysis through utilization of dynamical system Jacobians. Specifically, we investigate how the autoencoder architecture, latent space dimensionality, and, most critically, the training trajectory length ($n_t$)--parameters often selected heuristically--impact the accuracy and computational speedups achievable in the latent space. Through three distinct test cases, it is demonstrated how $n_t$ has a notable influence on the smoothness of latent space trajectories, enabling the use of larger time-steps in the latent space.}

This remainder of the paper is organized as follows. Section~\ref{Sec: Methodology} provides a description of the general autoencoder+neural ODE framework applied to a surrogate modeling task, including a distinction between different training methodologies and the proposed framework for time-scale analysis. The application of this framework to the Kuramoto-Sivashinsky (KS) equations is demonstrated in Section~\ref{Sec: KSE-eq}, with an analysis of the effect of network hyperparameters presented in Section~\ref{Sec: Effect of hyperparams}. Section~\ref{Sec: 1D Detonations} showcases the application of the framework to a highly advection-dominated test case of 1D detonation wave propgation.

\section{Methodology} \label{Sec: Methodology}
\subsection{Surrogate Modeling Task}

The application scope of this work is tied to accelerating simulation of physical systems influenced by advection -- particularly those governed by fluid dynamics -- using data-based surrogate models. Such systems can be mathematically described by partial differential equations (PDEs), where a general-form PDE is given by
\begin{equation} \label{eq: Continuous PDE}
    \frac{\partial \mathbf{\mathcal{U}}}{\partial t}  = \nabla\cdot  \mathcal{F}(\mathbf{\mathcal{U}}) + \nabla \cdot G \mathcal{U} + \mathcal{S}(\mathbf{\mathcal{U}}). 
\end{equation}
The above equation describes the evolution of a vector of state variables, denoted $\mathbf{\mathcal{U}} = \left({u}_1, {u}_2,.... {u}_{N_e}\right)$, where $N_e$ is the number of space- and time-dependent transport variables. The functions $\mathcal{F}$, $\mathcal{G}$, and $\mathcal{S}$ represent non-linear, linear, and volumetric source term operators respectively. Demonstrations in this work leverage data produced by one-dimensional numerical solutions of two model PDEs for fluid dynamics in which advection, through the nonlinear operator $\cal F$, plays a critical role. The first is the Kuramoto-Sivashinsky (KS) equations (Sec.~\ref{Sec: KSE-eq}), where the state vector is interpreted as a velocity magnitude ($N_e=1$). Here, the operator $\cal F$ represents an advection term, $\cal G$ a diffusive term, and the volumetric source term $\cal S$ is omitted. The second is the compressible reacting Navier-Stokes (NS) equations (Sec.~\ref{Sec: 1D Detonations}), where the state vector is higher dimensional, consisting of fluid density, velocity, chemical energy, and species mass fractions. The operators $\cal F$ and $\cal G$ are physically comparable with those used in the KS equations (i.e., they capture effects of advection and diffusion, respectively), and the volumetric source term $\cal S$ is retained, as it represents the effect of chemical reactions on the flow field. 

Regardless of the configuration, numerical solutions of Equation~\ref{eq: Continuous PDE} are obtained through a method-of-lines approach, which relies on spatial discretization onto a finite-dimensional grid. Here, the time-evolution of $\mathcal{U}$ sampled at all spatial discretization points -- denoted as $\mathbf{u} \in \mathbb{R}^{N_u}$, where $N_u$ is the full-system dimensionality computed as number of grid points multiplied by the number of transport variables $N_e$ -- is provided by the solution to a deterministic and high-dimensional ordinary differential equation (ODE)

\begin{equation} \label{eq: Discrete ODE}
    \frac{d \mathbf{u}(t)}{d t} = \mathbf{F}(\mathbf{u}(t)), \quad \mathbf{u}(t=0) = \mathbf{u}_0.
\end{equation}

In Eq.~\ref{eq: Discrete ODE}, $\mathbf{F}(\mathbf{u})$ captures the instantaneous discretized system dynamics (a discrete representation of the operators in Eq.~(\ref{eq: Continuous PDE}), and for the PDEs described above, represents complex interactions between advection, diffusion, and (if present) reaction contributions. Given an initial condition ${\bf u}_0$, Eq.~(\ref{eq: Discrete ODE}) can be solved to some final integration time as a system of ODEs using a proper time-integration scheme. Solution of the ODE for a given initial condition generates a trajectory of time-ordered snapshots of the high-dimensional system state variable, which serve as the training data for the data-driven modeling strategies described in the sections below. Note that this data can be produced either from explicit solutions of Eq.~\ref{eq: Discrete ODE} if the analytic PDE form is known (using time-integration schemes), or through real-world observations of the system in question if the PDEs are unknown or intractable to solve accurately (e.g., using high-speed or laser-based imaging tools). 

The ODE in Eq.~\ref{eq: Discrete ODE} is interpreted here as a "ground-truth" representation of the continuous PDE counterpart, meaning the grid is assumed to be resolved enough to properly capture the contribution of all spatiotemporal scales in the physical operators, resulting in a high-dimensional state vector ${\bf u}$ that corresponds to a physical space representation of the state. The motivation for surrogate modeling is that the ground-truth ODE in Eq.~\ref{eq: Discrete ODE} is computationally expensive to solve for realistic applications that require fast (near real-time) flow field predictions, such as full-scale design optimization~\cite{Amsallem2009, lieu2007adaptation, lieu2006reduced} or model-predictive control~\cite{mathelin2009robust, hovland2008explicit}. 

As such, the modeling goal is to identify an alternative surrogate ODE representation that expedites the solution to the ground-truth ODE without sacrificing predictive accuracy. The surrogate ODE is denoted
\begin{equation} \label{eq: Latent ODE}
    \frac{d {\mathbf{w}}(t)}{d t} = \mathbf{G}({\mathbf{w}}(t)), \quad {\mathbf{w}}(t_0) = \phi({\mathbf{u}(t_0)}).
\end{equation}

\begin{figure}
  \centering
  \includegraphics[width=1.0\textwidth]{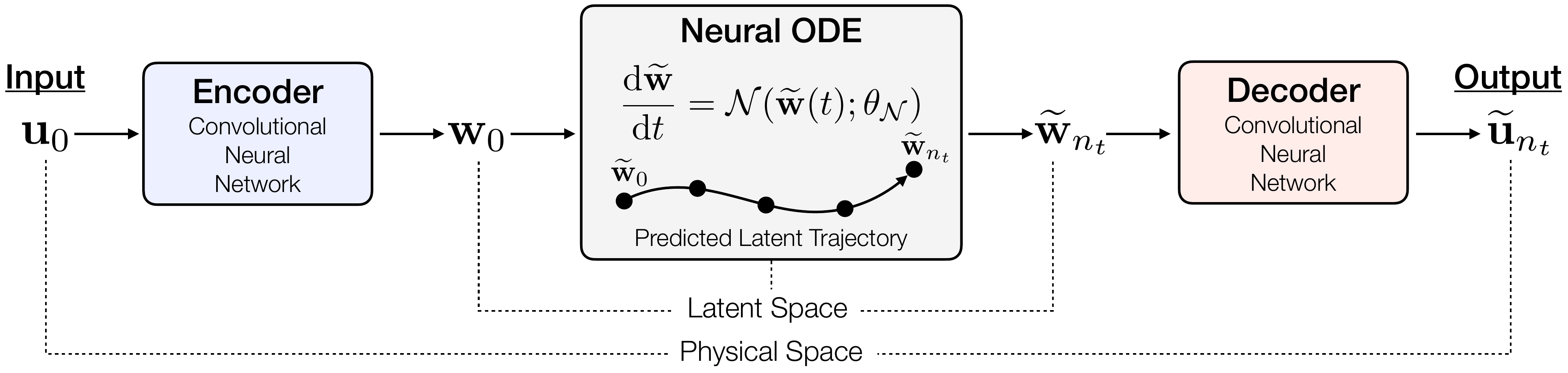}
  \caption{Combined autoencoder and neural ODE framework. Latent dynamics are modeled by a neural ODE, with movement between latent and physical spaces facilitated by an autoencoder. Neural ODE details are provided in Sec.~\ref{Sec: node}, and autoencoder details are provided in Sec.~\ref{Sec: autoencoder}}
  \label{fig:autoEncnODEschem}
\end{figure}

Instead of the full state variable ${\bf u}$, the surrogate ODE in Eq.~\ref{eq: Latent ODE} models the dynamics of a so-called \textit{latent} state vector ${\mathbf{w}} \in \mathbb{R}^{n_w}$, where $N_w \ll N_u$. The formulation of Eq.~\ref{eq: Latent ODE} highlights two key ingredients required to construct a data-based surrogate model, as outlined in Sec.~\ref{Sec: Intro}: (1) the function $\phi$, which is an instantaneous mapping function that transforms the original (physical space) state variable at a given time instant to its corresponding latent representation in a reduced space, and (2) the function ${\bf G}$, which is the latent dynamical system (it provides the dynamics in latent space). In this study, neural networks are used to provide functional forms for both components, the parameters of which are recovered in a training stage using an ensemble of trajectories from pre-computed solutions of the ground-truth ODE. More specifically, a neural ODE strategy is used to model the latent dynamics (described in Sec.~\ref{Sec: node}), while a convolutional autoencoder strategy is used to generate mappings to and from the reduced latent space (described in Sec.~\ref{Sec: autoencoder}). The overall approach is shown in Fig.~\ref{fig:autoEncnODEschem}

It should be noted that accelerated evaluations over the full system using a combined autoencoder and neural ODE strategy can be achieved through both a reduction in system dimensionality through the latent variables (the surrogate ODE operates on a lower-dimensional representation, which facilitates faster evaluations of instantaneous rates), and also an increase in minimum system timescales in the latent space inherent to the functional form of ${\bf G}$ \cite{Vijay23} (larger limiting timescales in ${\bf G}$ imply elimination of prohibitive timescales in ${\bf F}$, which in turn allows for larger time-steps to be utilized in the simulation procedure). More specifically, previous work has empirically observed how employing a convolutional autoencoder yields smooth trajectories in the latent space for chaotic PDE-based systems~\cite{Wan2023}. To this end, within the scope of data-based surrogate modeling, the primary emphasis of this work is to (a) provide a rigorous quantitative analysis on how timescale elimination is achieved in the latent space, (b) identify the role played by key neural ODE training hyper-parameters on this timescale elimination, and (c) understand the manner in which coupling between autoencoder and neural ODE components contributes to forecasting accuracy and latent timescale sensitivity.

Methodology for the components that facilitate this study is provided below. This includes details on neural ODE based modeling (Sec.~\ref{Sec: node}), integrating autoencoders into the neural ODE framework (Sec.~\ref{Sec: autoencoder}), description of training strategies (Sec.~\ref{Sec: Coupled vs Decoupled}), model evaluation metrics (Sec.~\ref{Sec: evaluation metrics}), and the extraction procedure for dynamical system timescales (Sec.~\ref{Sec: Dyn sys time-scales}).

\subsection{Neural ODEs for Latent Dynamics}
\label{Sec: node}

It is assumed that "ground-truth" data represented by an ensemble of latent space trajectories -- generated by application of an autoencoder to a corresponding set of full-order system trajectories (to be described in Sec.~\ref{Sec: Coupled vs Decoupled}) -- are available. These trajectories are assumed to be sampled at a fixed discrete time interval $\Delta t$. As a result, one such ground-truth latent trajectory is given by the time-ordered set
\begin{equation}
    \label{eq:ground_truth_latent}
    {\mathcal{T}}_{w,i} = [{\bf w}(t_0), {\bf w}(t_0 + 1 \Delta t), {\bf w}(t_0 + 2 \Delta t), \ldots, {\bf w}(t_0 + n_t \Delta t)], \quad i = 1, \ldots, N_T, 
\end{equation}
where $n_t$ corresponds to the so-called \textit{training trajectory length}, and $N_T$ represents the total number of ground-truth latent trajectories obtained at a respective $n_t$. In this work, the ground-truth \textit{latent} trajectories in Eq.~\ref{eq:ground_truth_latent} are obtained from the action of a convolutional neural network (CNN) based encoder on the corresponding full-order trajectory (i.e., ${\cal T}_{w,i} \leftarrow \text{Encode}({\cal T}_{u,i})$, where ${\cal T}_{u,i}$ contains the full-order snapshots). 

\begin{figure}
  \centering
  \includegraphics[width=\textwidth]{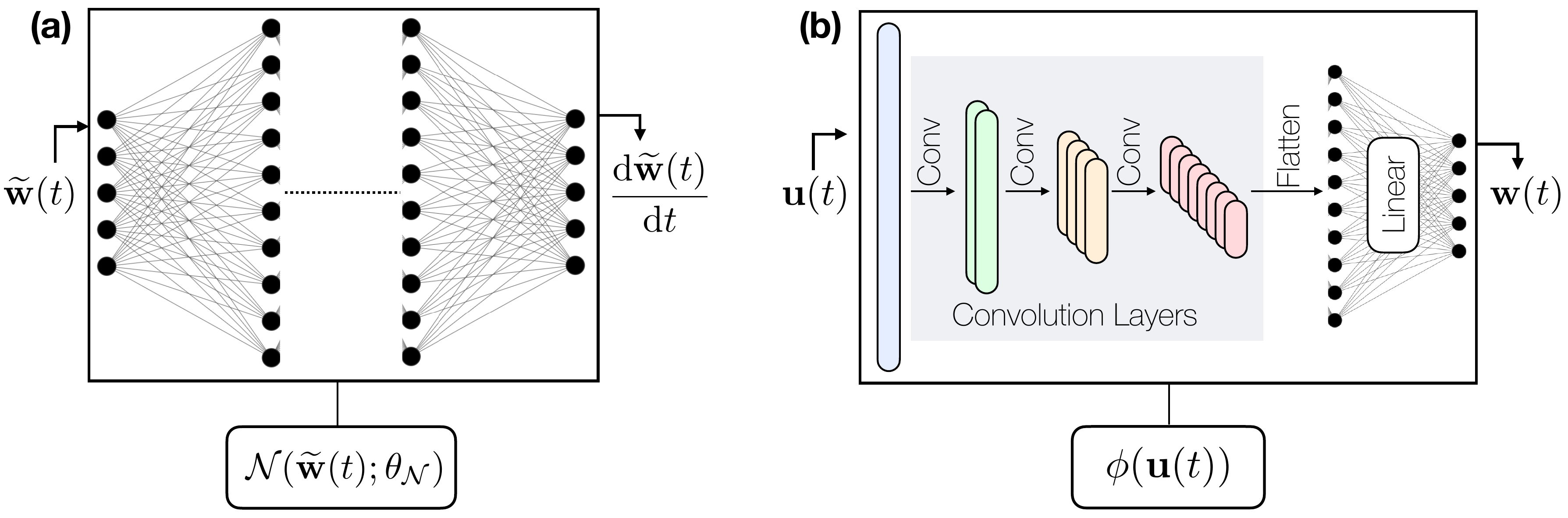}
  \caption{\textbf{(a)} Schematic of neural ODE operation. The architecture of ${\cal N}$ is a feed-forward neural network with four hidden layers and an ELU activation functions. The number of hidden units is 120 neurons. \textbf{(b)} Schematic of encoder operation. Encoder contains a sequence of 1D convolutional layers with batch normalization and an ELU activation function, in which the physical space input ${\bf u}(t)$ is progressively down-sampled in space. The spatial component is down-sampled by a factor of two while doubling the number of channels, upon encountering a flattening operation and linear layer to produce the latent space projection ${\bf w}(t)$. Although not shown, the decoder $\psi$ is a mirrored version of the encoder, with transpose convolution layers replacing convolution layers. Both architectures are implemented in PyTorch \cite{pytorch}.}
  \label{fig: nODE_autoenc_schem}
\end{figure}

Neural ODEs leverage the above trajectory data to learn a continuous-time model for the unknown latent dynamical system that governs the evolution of ${\bf w}$ \cite{chen2018neural}. The starting point is to cast the functional form for the latent dynamics (the right-hand-side of Eq.~\ref{eq: Discrete ODE}) as a neural network, resulting in the neural ODE
\begin{equation} \label{eq: nODE_approx}
    \frac{d \widetilde{\mathbf{w}}(t) }{d t} = \mathcal{N}(\widetilde{\mathbf{w}}(t); \theta_{\cal N}), \quad \widetilde{\mathbf{w}}(t_0) = {\bf w}(t_0). 
\end{equation}
In Eq.~\ref{eq: nODE_approx}, $\mathcal{N}$ is a neural network characterized by parameter set $\theta_{\cal N}$. Figure~\ref{fig: nODE_autoenc_schem} (left) provides a description for the architecture of ${\cal N}$ used in this work. Given an initial condition ${\mathbf{w}}_0$, a black box time integrator can be used to find the latent state $\widetilde{\mathbf{w}}(t)$ at any time $t$. In this work, an explicit Euler time-integration scheme with a constant time-step ($\Delta t$) is used. As such, given the above neural ODE, a \textit{predicted} latent trajectory analogous to the ground-truth trajectory of Eq.~\ref{eq:ground_truth_latent}, starting at the same initial condition, can be represented as
\begin{equation}
    \label{eq:predicted_latent}
    \widetilde{\mathcal{T}}_{w,i} = [\widetilde{\bf w}(t_0), \widetilde{\bf w}(t_0 + 1 \Delta t), \widetilde{\bf w}(t_0 + 2 \Delta t), \ldots, \widetilde{\bf w}(t_0 + n_t \Delta t)], \quad i = 1, \ldots, N_T. 
\end{equation}

% In Eq.~\ref{eq:node_loss}, $\Tilde{\hat{\mathbf{u}}}_{n_t}^i$ is the $i^{th}$ sub-trajectory, consisting of $n_t$ discrete points, predicted by integrating the neural ODE in time,
% \begin{multline} \label{eq: predTraj}
%         \Tilde{\hat{\mathbf{u}}}_{n_t}^i =  [ \left( \hat{\mathbf{u}}(i n_t \Delta t) +  \int^{i n_t + \Delta t}_{i n_t}  \mathcal{N}(\hat{\mathbf{u}}, t, \theta_N) \,dt \right), \quad \left( \hat{\mathbf{u}}(i n_t \Delta t) +  \int^{i n_t + 2\Delta t}_{i n_t}  \mathcal{N}(\hat{\mathbf{u}}, t, \theta_N) \,dt \right) ,.... \quad \\ \left( \hat{\mathbf{u}}(i n_t \Delta t) +  \int^{ i n_t + i n_t \Delta t}_{i n_t}  \mathcal{N}(\hat{\mathbf{u}}, t, \theta_N) \,dt \right) ]  ,
% \end{multline}
% and $\hat{\mathbf{u}}_{n_t}^i$ is the corresponding truth data trajectory, 
% \begin{equation}
%     \hat{\mathbf{u}}_{n_t}^i = \left[ \hat{\mathbf{u}}(i n_t \Delta  + \Delta t), \quad \hat{\mathbf{u}}(i n_t \Delta  + 2 \Delta t),... \quad \hat{\mathbf{u}}(i n_t \Delta  + n_t \Delta t) \right].
% \end{equation} 
% The total number of trajectories is computed as,
% \begin{equation}
%     N_{tr} = \frac{L_{tr}}{n_t},
% \end{equation}
% where $L_{tr}$ is the total length of the available truth data trajectory. 

\begin{figure}
  \centering
  \includegraphics[width=0.6\textwidth]{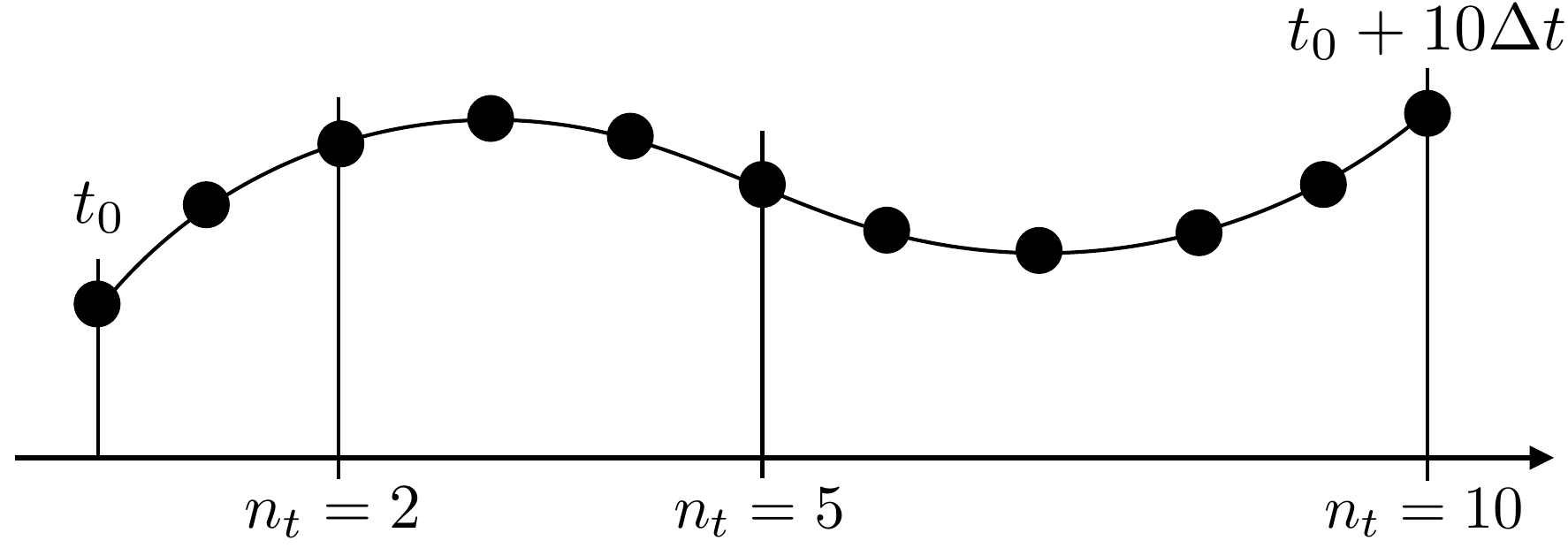}
  \caption{Schematic illustrating the interpretation of training trajectory length $n_t$, which controls how long a NODE prediction is rolled-out during training. Higher $n_t$ values during training allow for more dynamical information to be included in the training objective.}
  \label{fig:trajSpliSchem}
\end{figure}

To optimize the parameters $\theta_{\cal N}$ of the neural network $\cal N$, an objective function representing a mean-squared error between the ${\mathcal{T}}_{w,i}$ ground-truth latent trajectories and the $\widetilde{\mathcal{T}}_{w,i}$ predicted latent trajectories is minimized in a training stage. This objective function is given by
%(${n_{\hat{\mathbf{u}}}}$), chosen trajectory length ($n_t$), and the total number of sub-trajectories ($N_{tr}$)
\begin{equation}
    \label{eq:node_loss}
    \mathcal{L}_{\text{NODE}} = \left \langle \frac{1}{n_t} \sum_{j = 1}^{n_t} \big\lVert {\bf w}(t_0 + j\Delta t) - \widetilde{\bf w}(t_0 + j\Delta t) \big\rVert_2^2 \right \rangle, 
\end{equation}
where the angled brackets $\langle . \rangle$ represent an average over a batch of training set trajectories. The formulation in Eq.~\ref{eq:node_loss} reveals the primary advantage of the neural ODE formulation, in that the training approach is \textit{dynamics-informed}: the latent dynamical system is trained to minimize accumulated error in the latent trajectory over all $n_t$ time-steps. As such, it must be emphasized that $n_t$ -- the training trajectory length -- is a critical training parameter that captures the amount of dynamical information used to construct the latent dynamical system, and is typically chosen and fixed \textit{a-priori}. The implication is that the training procedure can be executed with different $n_t$ values from the same time-ordered data, resulting in neural ODEs with different predictive capabilities and stability properties. To illustrate this concept, the schematic in Fig.~\ref{fig:trajSpliSchem} shows how a single large ground-truth latent trajectory can be split into $N_T$ training trajectories, each of length $n_t$. A primary goal of this work is to rigorously study the effect of $n_t$ on both predictive accuracy and latent time-scales, in addition to proposing a pathway to choose an optimal value of $n_t$ based on the underlying physics of the problem. 

It should be noted that in the training stage, in order to calculate the derivatives necessary for optimization ( $\frac{\partial \mathcal{L}_{\text{node}}}{\partial \theta}$) two approaches can be used: one involves storing the entire rollout graph of the integrated ODE trajectory and propagating gradients backward through it, while the other entails solving a system of ODEs known as the adjoint equations backward in time. In this study, the latter method is used to ensure memory efficiency when dealing with long training trajectory lengths. This work leverages the \verb|torchdiffeq| library \cite{torchdiffeq} for neural ODE training routines.

% \begin{figure}[h!]
%   \centering
%   \includegraphics[width=1.0\textwidth]{Schematics/nODESchem.pdf}
%   \caption{Neural ODE Schematic}
%   \label{fig:nODESchem}
% \end{figure}
% Fig.~\ref{fig:nODESchem}. shows the architecture used for the neural ODE, it is a simple feed-forward neural network with four layers, and an ELU activation after each layer. The number of hidden units in each layer was chosen empirically, to balance accuracy and training times. 

\subsection{Dimensionality Reduction with Autoencoders} \label{Sec: autoencoder}

% Integrating autoencoders with neural ordinary differential equations (ODEs) offers a novel approach to constructing reduced-order models, leveraging autoencoders for mapping to and from the latent space. This coupled with the continuous-time latent dynamics modeled by neural ODEs, facilitates the efficient construction of reduced-order representations.

Autoencoders are a class of neural networks used for unsupervised learning, primarily in the domain of data compression and feature extraction. An autoencoder consists of an encoder and a decoder, working in concert to both achieve dimensionality reduction in the input high-dimensional data while preserving all of its salient information. 

As mentioned in Sec.~\ref{Sec: node}, to generate ground-truth latent trajectories in Eq.~\ref{eq: Latent ODE}, and to facilitate neural ODE inference in latent space, a neural network based encoder $\phi$ is used to project instantaneous full-order state samples ${\bf u}(t) \in \mathbb{R}^{N_u}$ to corresponding lower-dimensional latent representations ${\mathbf{w}}(t) \in \mathbb{R}^{N_{w}}$ such that $N_{w} \ll N_u$. The latent trajectories described in Eq.~\ref{eq:ground_truth_latent} are then recovered by applying the trained encoder $\phi$ to the corresponding high-fidelity ground-truth trajectories. 

The objective of the decoder $\psi$ is to then reconstruct the original state from this latent representation with minimal loss of information, i.e. ${\bf u}(t) \approx \widetilde{\bf u}(t) = \psi( \phi ( {\bf u} (t) ) )$. Inspired by recent work in data-based reduced order modeling \cite{romit_cnn_advection,Wan2023,lee2020model}, this work leverages convolutional neural networks as the backbone for encoder and decoder architectures. Figure~\ref{fig: nODE_autoenc_schem}(right) illustrates the encoder architecture; the decoder is a mirrored version of the encoder, which transpose convolution layers replacing convolution layers. Alongside the architectural configuration, the critical parameter in any autoencoder is the latent space dimensionality $N_{w}$, which effectively controls the trade-off between dimensionality reduction and reconstruction accuracy.

The encoder and decoder are trained by minimizing the loss function 
\begin{equation} \label{eq: Autoenc_loss}
    \mathcal{L}_{\text{AE}} =  \left\langle \big\lVert {\bf u}(t) - \psi( \phi ( {\bf u} (t) ) \big\rVert_2^2 \right\rangle
\end{equation}
that characterizes the mismatch between the reconstructed and original states. The angled brackets in the above equation denote an average over all ${\bf u}(t)$ instantaneous target snapshots in the training set. Note that the time-ordered quality of the data is crucial when training the neural ODE in latent space, but is unimportant when training the autoencoder, since both encoder and decoder are instantaneous mapping functions.

% \begin{figure}[h!]
%   \centering
%   \includegraphics[width=1.0\textwidth]{Schematics/Group Meeting_nODE.pdf}
%   \caption{Encoder Schematic}
%   \label{fig:AutoencoderSchem}
% \end{figure}

% Fig~\ref{fig:AutoencoderSchem}. details the encoder architecture used in this work, which contains a sequence of convolutional layers with batch normalization and a ELU activation function, in which the input is progressively downsampled in space. The spatial component is downsampled by a factor of two while doubling the number of channels, thus maintaining the same amount of information throughout the layers. The decoder is an exact transpose of the encoder. 

% Figure~\ref{fig:autoEncnODEschem} details how the autoencoder are neural ODE are combined to predict trajectories in the state space $\Tilde{\mathbf{u}}^i_{n_t} =  \psi(\mathbf{\Tilde{\hat{u}}}_{n_t}^i)$.

% The neural ODE is a feed-forward neural network with four hidden layers, and an ELU activation after each layer. The number of hidden units in each layer was chosen empirically to balance accuracy and training time. The autoencoder contains a sequence of convolutional layers with batch normalization and an ELU activation function, in which the input is progressively downsampled in space. The spatial component is downsampled by a factor of two while doubling the number of channels, thus maintaining the same amount of information throughout the layers. The decoder is an exact transpose of the encoder.

\subsection{Coupled versus Decoupled Training}
\label{Sec: Coupled vs Decoupled}
There are two distinct ways of training the combined autoencoder and neural ODE based model in Fig.~\ref{fig:autoEncnODEschem}. The first approach, termed \textit{decoupled training}, entails training the autoencoder and neural ODE separately, treating them as two distinct optimization problems. The second approach, termed \textit{coupled training}, involves simultaneous training of the autoencoder and the neural ODE. The details of the differences between the two is explained below. The effect of these training methodologies on both predictive accuracy and latent time-scales is explored in Sec.~\ref{Sec: Numerical Experiments}. 

\textbf{Decoupled Training:} The initial step involves training the autoencoder with snapshots of the state vector at various parameter instances, optimizing the autoencoder loss $\mathcal{L}_{\text{AE}}$ using full-order trajectory data ${\cal T}_{u,i}$. Subsequently, the trained encoder is employed to produce the ground-truth latent trajectories ${\cal T}_{w,i}$. Following this, the neural ODE is trained by minimizing the neural ODE loss $\mathcal{L}_{\text{NODE}}$ within the latent space, as described in Eq.~\ref{eq:node_loss}. 

\textbf{Coupled Training:} In this approach, the autoencoder and neural ODE are trained concurrently using a single optimization problem in the full state-space. The loss that is minimized is given by,

\begin{equation} \label{eq: coupled training loss}
    \mathcal{L}_{\text{coupled}} = \mathcal{L}_{1} + \mathcal{L}_{2},
\end{equation}
where $\mathcal{L}_{1}$ and $\mathcal{L}_{2}$ are defined as
\begin{equation} \label{eq: coupled_loss}
    \mathcal{L}_{1} = \left \langle \frac{1}{n_t}\sum_{j=1}^{n_t} \big\lVert \psi(\widetilde{\bf w}(t + j \Delta t)) - \mathbf{u}(t + j \Delta t) \big\rVert_2^{2} \right \rangle 
    \quad \text{and} \quad
    \mathcal{L}_{2} =  {\cal L}_{\text{AE}}
\end{equation}

respectively. The quantity ${\cal L}_1$ represents the error due to the neural ODE-predicted trajectory in physical space, and $\mathcal{L}_{2}$ is added to improve the instantaneous projection capabilities of the autoencoder. The implication of the coupled approach is that the autoencoder parameters are also informed of the intrinsic system dynamics in the training stage, which is not the case in the de-coupled approach.

% Figure~\ref{fig:CoupVsDecoup} provides a schematic comparing the coupled and decoupled training approaches. 

% The effect of this loss term on the reconstruction accuracy of the trained autoencoder is explored further in Section {\color{red} add section}.

% \begin{figure}[h!]
%   \centering
%   \includegraphics[width=1.0\textwidth]{Schematics/CoupledVsDecoupledSchem.pdf}
%   \caption{Coupled vs Decoupled training.}
%   \label{fig:CoupVsDecoup}
% \end{figure}

\subsection{Evaluation Metrics}
\label{Sec: evaluation metrics}
To compare the predictive accuracy of different combined autoencoder neural ODE models, two testing metrics are used. The first metric, called the single-step prediction error ($\mathcal{L}_{\text{SS}}$) is defined as
\begin{equation} \label{eq: SSR_error}
    \mathcal{L}_{\text{SS}} = \left\langle || \psi( \widetilde{\bf w} (t + \Delta t)) - {\bf u}(t + \Delta t) || _2^{2} \right\rangle,
\end{equation}

where the angled brackets represent an average over all $(t, t+\Delta t)$ testing snapshot pairs. The single-step prediction error $\mathcal{L}_{\text{SS}}$ can therefore be used to provide a \textit{a-priori} measure of neural ODE predictive accuracy along a trajectory, in that the effect of error accumulation through time is discarded.

The second metric is the rollout error, and is given by
% nvolving computing the mismatch error over a chosen trajectory length $n_{\text{RO}}$ 
\begin{equation}  \label{eq: RO_error}
    \mathcal{L}_{\text{RO}} = \left \langle \frac{1}{n_{\text{RO}}}\sum_{j=1}^{n_{\text{RO}}} \big\lVert \psi(\widetilde{\bf w}(t + j \Delta t)) - \mathbf{u}(t + j \Delta t) \big\rVert_2^{2} \right \rangle.
\end{equation}
The rollout error metric in Eq.~\ref{eq: RO_error} represents an \textit{a-posteriori} evaluation of the neural ODE in physical space, and is nearly identical to ${\cal L}_1$ in Eq.~\ref{eq: coupled_loss}. The difference is that the trajectory error in Eq.~\ref{eq: RO_error} above is computed for an \textit{evaluation} trajectory of size $n_{\text{RO}}$ instead of a \textit{training} trajectory of size $n_t$. In other words, a neural ODE trained using trajectory lengths $n_t$ can be evaluated in inference to produce trajectory lengths of different sizes. In the formulation above, $n_{\text{RO}} = 1$ recovers the single-step prediction error in Eq.~\ref{eq: SSR_error}. The corresponding Relative Absolute Error (RAE) version of $ \mathcal{L}_{\text{SS}}$ is given by,
\begin{equation} \label{eq: RSS Loss }
    \mathcal{R}_{\text{SS}} = \Biggl \langle \frac{ \left| \psi(\mathbf{\widetilde{{w}}}(t+\Delta t)) - \mathbf{u}(t+\Delta t) \right| }{\left| \mathbf{u}(t+\Delta t) \right|} \Biggr \rangle,
\end{equation} and the same for $\mathcal{L}_{\text{RO}}$ is given by,
\begin{equation} \label{eq: RRO Loss }
    \mathcal{R}_{\text{RO}} = \Biggl \langle \sum^{n_{RO}}_{j=1} \frac{ \left| \psi(\mathbf{\widetilde{{w}}}(t + j \Delta t)) - \mathbf{u}(t + j \Delta t) \right| }{\left|  \mathbf{u}(t + j \Delta t) \right|} \Biggr \rangle.
\end{equation}
As such, in the demonstrations in Sec.~\ref{Sec: Numerical Experiments}, one of the objectives is to isolate the effect of various neural ODEs produced with different training trajectory lengths $n_t$ on the above evaluation metrics. 
% where $\mathbf{\Tilde{\hat{u}}}_{n_{\text{RO}}}^i$ is a sub-trajectory of length $n_{\text{RO}}$ given by Eq.(\ref{eq: predTraj}). The total number of trajectories in this case is denoted as $N_{tr} = L_{tr}/n_{\text{RO}}$. The corresponding truth data trajectory is given by,
% \begin{equation}
%     {\mathbf{u}}_{n_{\text{RO}}}^i = \left[ {\mathbf{u}}(i n_{\text{RO}} \Delta  + \Delta t), \quad {\mathbf{u}}(i n_{\text{RO}} \Delta  + 2 \Delta t),... \quad {\mathbf{u}}(i n_{\text{RO}} \Delta  + n_t \Delta t) \right].
% \end{equation}

\subsection{Dynamical System Timescales} \label{Sec: Dyn sys time-scales}

Eigenvalue analysis has been used to analyse the time-scales of dynamical systems~\cite{Maas1992, Mease2003, Valorani2001}. Training neural ODEs on long trajectories is a challenging problem which requires a lot of fine-tuning. The choice of the training trajectory length $n_t$ is often based on heuristics and does not have any theoretical justification. The complexity of the training loss function grows with the increase in $n_t$, and a neural ODE trained on a small $n_t$ may not capture the relevant dynamics of the system. Thus, choosing an optimal $n_t$ is an open research problem. In this work, we employ an eigenvalue based time-scale analysis framework to rigorously study the effect of $n_t$ on accuracy and time-scale reduction in the latent space. We employ the same framework to study the effect of certain network architecture based hyperparamaters on time-scale reduction in the latent space. For a dynamical system that is governed by an ODE given by Eq.~(\ref{eq: Discrete ODE}), the inverse of the largest eigenvalue of the right-hand side Jacobian,
\begin{equation} \label{eq: fullSys_tlim}
   t_{lim}(\mathbf{F}, t) = \frac{1}{max ( | eig(\frac{\partial \mathbf{F}(\mathbf{U}, t)}{\partial \mathbf{U}}) | )},
\end{equation}
gives a measure of the fastest evolving time-scales and can give an indication of the largest time-step that can be used to evolve the system using an explicit time-integration scheme. A higher value of $t_{lim}$ in the latent space indicates smoother latent space trajectories and in-turn larger possible time-steps. The inverse of the smallest eigenvalue on the other hand,
\begin{equation} \label{eq: fullSys_tmax}
   t_{max}(\mathbf{F}, t) = \frac{1}{min ( | eig(\frac{\partial \mathbf{F}(\mathbf{U}, t)}{\partial \mathbf{U}}) | )},
\end{equation}
gives a measure of the slowest evolving time-scales of the system. In this work, we use this metric to find a correlation between $n_t$ and the predictive accuracy of the combined autoencoder neural ODE framework.

\section{Numerical Experiments} \label{Sec: Numerical Experiments}

The combined autoencoder neural ODE framework is tested in two numerical experiments: the Kuramoto–Sivashinsky (KSE) equations and a one-dimensional detonation problem. The KS equation serves as a test bed to discern the impact of various training hyperparameters on time-scale reduction in the latent space. Eigenvalue analysis, as explained in the previous section, quantifies the fastest and slowest time scales in the latent space. For a latent dynamics model approximated by a neural ODE, described by Eq.(\ref{eq: nODE_approx}), the fastest and slowest evolving time-scales in the latent space are given by,
\begin{equation} \label{eq: nODE tlim tmax}
    t_{\text{lim}}(t) = \frac{1}{\text{max} ( | \text{eig}(\frac{\partial \mathcal{N}(\widetilde{\mathbf{w}}(t); \theta_{\cal N})}{\partial \widetilde{\mathbf{w}}}) | )} \quad \text{and} \quad t_{\text{max}}(t) = \frac{1}{\text{min} ( | \text{eig}(\frac{\partial \mathcal{N}(\widetilde{\mathbf{w}}(t); \theta_{\cal N})}{\partial \widetilde{\mathbf{w}}}) | )},
\end{equation}

To assess the impact of the number of convolutional layers, latent dimensions in the autoencoder, and the training methodology, networks with varied configurations for each parameter (while keeping the other two fixed) are trained. The resulting $t_{max}(\textit{nODE}, t)$ profiles are then compared to the full system time-scales provided by Eq.(\ref{eq: fullSys_tlim}). Furthermore, the influence of the training trajectory length $n_t$ on both time-scale reduction and predictive accuracy is considered for both numerical experiments.

\subsection{KS Equation} \label{Sec: KSE-eq}

The KS equation is solved using \verb|jax-cfd| \cite{Kochkov_2021}, which allows differentiation through the full-order (physical space) dynamical system \( d{\bf u}/dt=\mathbf{F(u)} \) , which facilitates Jacobian computation \( \frac{\partial \mathbf{F}}{\partial \mathbf{u}} \) for extracting the corresponding \textit{true} physical space time-scales facilitating comparison with the time-scales in the latent space.
The KS equation is defined by the partial differential equation,
\begin{equation}
    \frac{\partial u}{\partial t} + u\frac{\partial u}{\partial x} + \nu\frac{\partial^2 u}{\partial x^2} - \nu\frac{\partial^4 u}{\partial x^4} = 0 \quad \text{in } [0, L] \times \mathbb{R}^+, 
\end{equation}
with periodic boundary conditions, where \(L = 64\). The initial conditions are specified as,
\begin{equation}
    u_0(x) = \sum_{k=1}^3 \sum_{j=1}^{nc} a_j \sin(\omega_j \cdot x + \phi_j), 
\end{equation}

where \(\omega_j\) is randomly chosen from \(\{\pi/L, 2\pi/L, 3\pi/L\}\), \(a_j\) is sampled from a uniform distribution in \([-0.5, 0.5]\), and the phase \(\phi_j\) follows a uniform distribution in \([0, 2\pi]\). The parameter \(n_c = 30\) governs the number of modes in the initial conditions. $\nu=0.01$ and other mentioned parameters are chosen to match previous work~\cite{Wan2023, dresdner2023learning, Bar_Sinai_2019}. 

The KS equation is solved on a uniform domain of size 1024 using a pseudospectral discretization. A pseudospectral solver equations utilizes spectral methods for solving partial differential equations (PDEs). The spatial domain is discretized using Chebyshev collocation points, which provide accurate approximations for functions with minimal numerical dispersion. The solver employs a Fourier transform to represent the temporal derivatives in the KS equation, converting the PDE into a system of ordinary differential equations (ODEs) in the spectral domain. Time integration is performed using an explicit Euler method and the integrated trajectories are converted back to physical space using an inverse Fourier transform. This pseudospectral approach allows for efficient and accurate simulation of the spatiotemporal dynamics governed by the KS equation. The psuedospectral solver implemented jax-cfd is used to generate the data. The training samples are generated by assembling 25 trajectories, from different initial conditions, each comprising of 1500 time-steps ($\Delta t = 1.95e^{-3}$) and the trained autoencoder neural ODE framework is tested on unseen initial conditions. 

\begin{figure}
  \centering
  \includegraphics[width=0.66\textwidth]{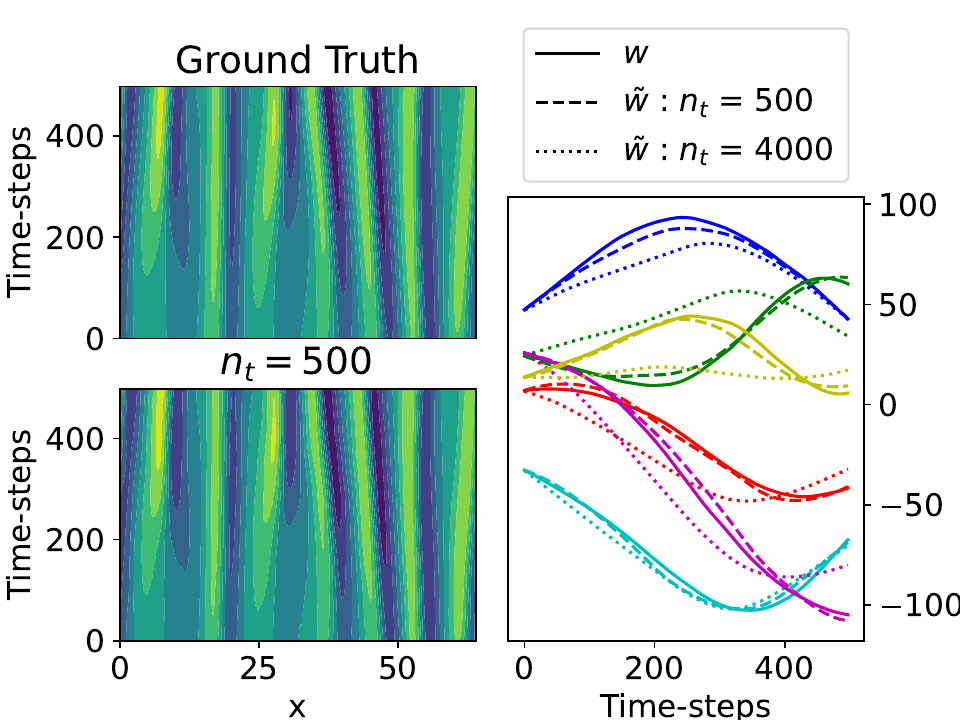}
  \caption{(Left) Comparison of the ground truth and rollout predicted fields (latent dimension of 25 and 4 convolutional layers and $n_t=500$) for an unseen initial condition. (Right) Corresponding latent space trajectories for $n_t=500$ (dashed) and $n_t=4000$ (dotted), compared to ground truth (solid). A subset of the 25 latent trajectories are shown here for visual clarity.}
  \label{fig:KSE Latent traj}
\end{figure}

Figure~\ref{fig:KSE Latent traj} compares the ground truth field for $u$, with the rollout field ($n_{RO}=500$) predicted by the autoencoder neural ODE framework trained in a decoupled manner using $n_t=500$, for an unseen initial condition. The predictions closely match the ground truth data. The figure also illustrates the corresponding latent space trajectories for $n_t=500$ and $n_t=4000$, both of which appear smoother compared to the state space fields. Although the presence of smooth latent space trajectories has been empirically observed~\cite{Wan2023}, it has not been quantitatively studied.  It can be seen that the latent space trajectories predicted by the neural ODE with $n_t = 500$ more closely matches the ground truth trajectories as compared to the neural ODE with $n_t = 4000$. These noticeable differences in accuracy suggest that the relationship between increasing $n_t$ and predictive strength is not straightforward, as discussed later. Additionally, latent space trajectories are smoother than state space fields, with $n_t = 4000$ trajectories being smoother than $n_t = 500$.

% \begin{figure}
%   \centering
%   \includegraphics[width=0.49\textwidth]{Results/latTraj.pdf}
%    \includegraphics[width=0.49\textwidth]{Results/fieldPlots.pdf}
%   \caption{(Left) Comparison of the ground truth and predicted fields for an unseen initial condition (Right) Corresponding latent space trajectories.}
%   \label{fig:KSE Latent traj}
% \end{figure}

\subsubsection{Effect of network hyperparameters} \label{Sec: Effect of hyperparams}

In most applications of the combined autoencoder and neural ODE approach for dynamical systems, hyperparameters related to the network architecture are often chosen heuristically. In this section, the effect of two hyperparameters on the time-scale reduction is quantified. 

\textbf{Number of convolutional layers}: The convolution operation is conceptualized as a localized filtering operation, suggesting that incorporating more convolutional layers between the state space and the latent space should ideally yield smoother latent space trajectories and, consequently, greater time-scale reduction in the latent space.

\begin{figure}
  \centering
  \includegraphics[width=0.49\textwidth]{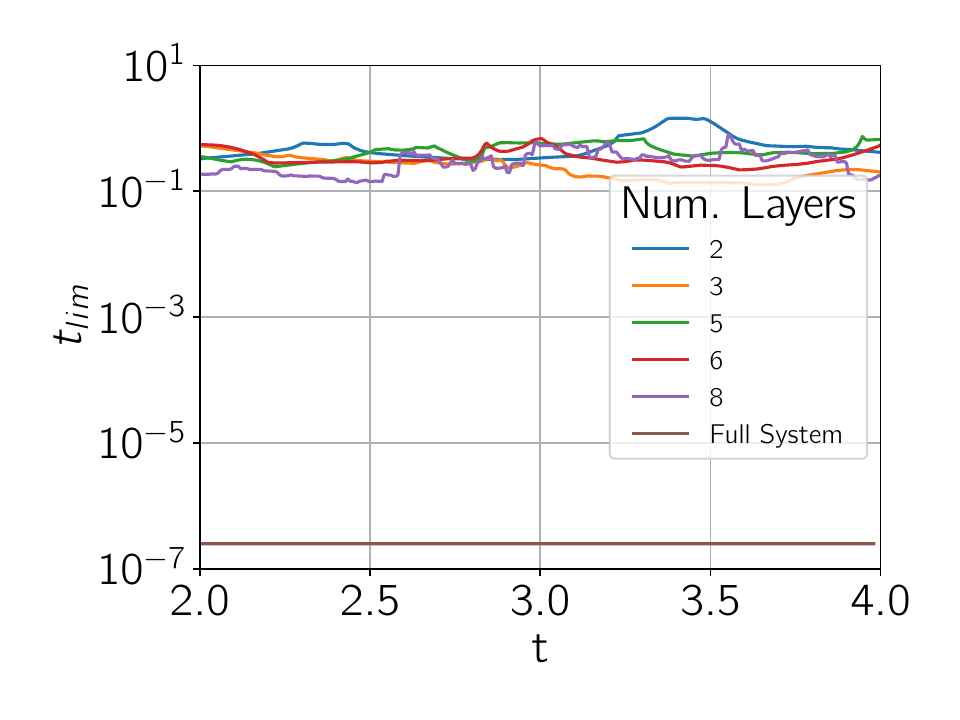}
\includegraphics[width=0.49\textwidth]{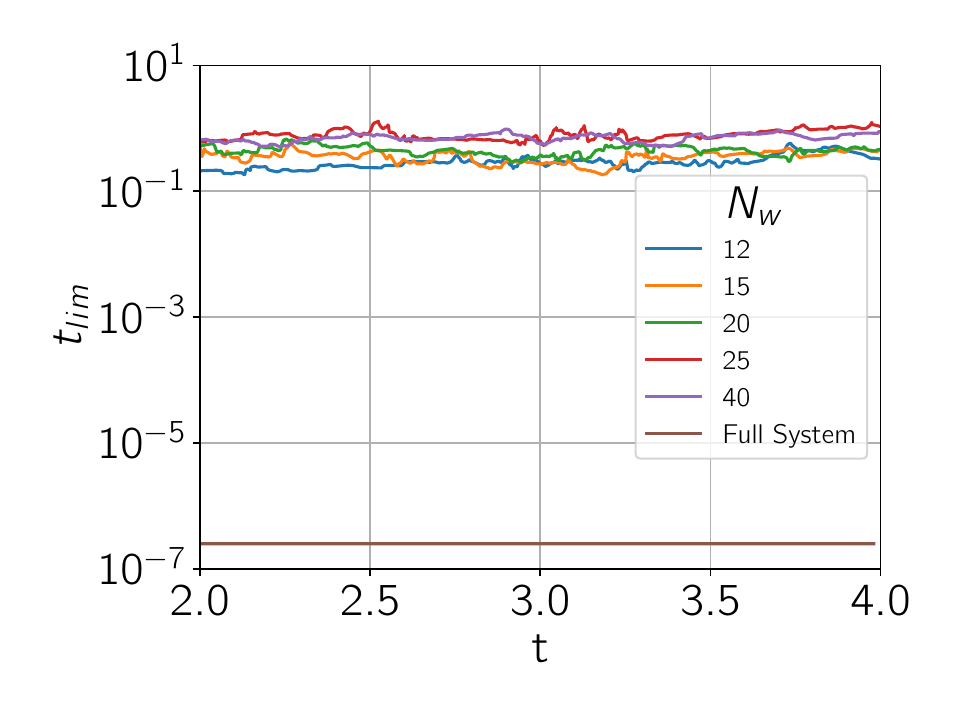}
  \caption{Limiting time-scale in the latent space $t_{lim}$ (Eq.~\ref{eq: nODE tlim tmax}) as a function of time for neural ODEs trained with varying (left) different number of autoencoder convolutional layers (using $n_t=500$ and latent dimension of 25) and (right) latent dimensionality $N_{w}$ (using 4 convolution layers and $n_t=500$). }
  \label{fig:tlim_convLayers}
\end{figure}

However, Fig.~\ref{fig:tlim_convLayers} illustrates the limiting time-scale in the latent space, denoted as $t_{lim}(\textit{nODE}, t)$, as a function of time for autoencoders with varying numbers of convolutional layers. Contrary to expectations, the number of convolutional layers does not exert a significant influence on the time-scales of the latent variables. In Fig.~\ref{fig:tlim_convLayers}, the line corresponding to the 'full-system' is generated by plotting the inverse of the largest eigenvalue of the full KSE-equations right-hand side Jacobian as a function of time. It can be seen that all the autoencoders in general significantly reduce the limiting time-scales of the system, by roughly six orders of magnitude.

\textbf{Latent space dimensionality}: Figure~\ref{fig:tlim_convLayers} also depicts the limiting time-scale in the latent space, denoted as $t_{lim}(\textit{nODE}, t)$, as a function of time for autoencoders with different numbers of latent dimensions. The observation from the figure is that the size of the latent space does not have a substantial impact on the limiting time-scales in the latent space.

% \begin{figure}
%   \centering
%   \caption{Limiting time-scale in the latent space $t_{lim}(\textit{nODE}, t)$ as a function of time for autoencoders with different latent dimension sizes ($n_{\hat{u}}$).}
%   \label{fig:tlim_latDim}
% \end{figure}

\subsubsection{Effect of training methodology}

Literature on the application of the combined autoencoder neural ODE for surrogate modeling lacks a rigorous comparison between the coupled and decoupled training approaches described in Section~\ref{Sec: Coupled vs Decoupled}. This section studies the effect of the training methodology on predictive accuracy along with the effect of adding the $\mathcal{L}_{2}$ loss term on the accuracy of the autoencoder.

\textbf{Loss Terms}: The effect of adding $\mathcal{L}_{2}$ loss term to the coupled training approach described in Section ~\ref{Sec: Coupled vs Decoupled}, is studied by comparing autencoder reconstruction loss $\mathcal{L}_A$ of coupled autoencoders trained with and without the $\mathcal{L}_{2}$ loss term added during training. 

\begin{figure}
  \centering
  \includegraphics[width=0.49\textwidth]{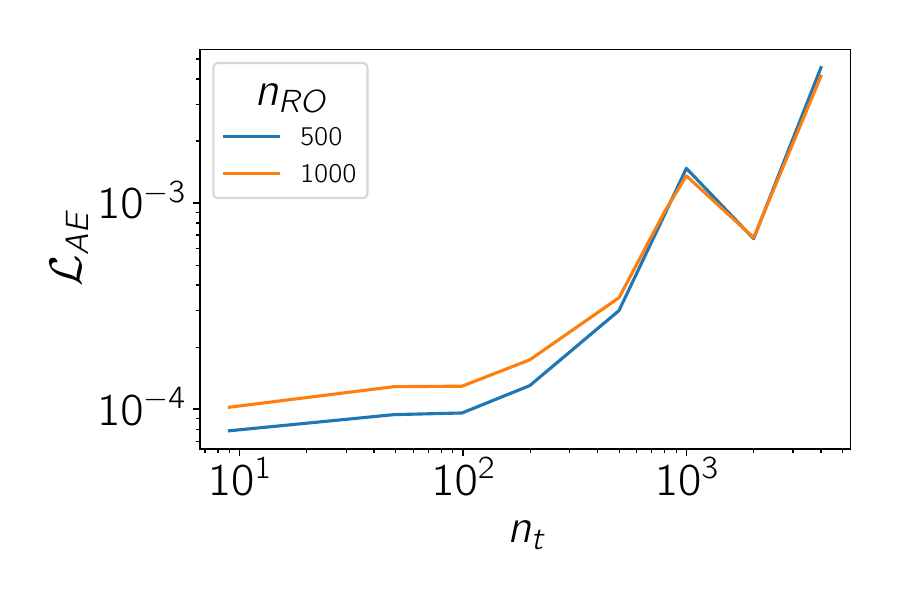}
  \includegraphics[width=0.49\textwidth]{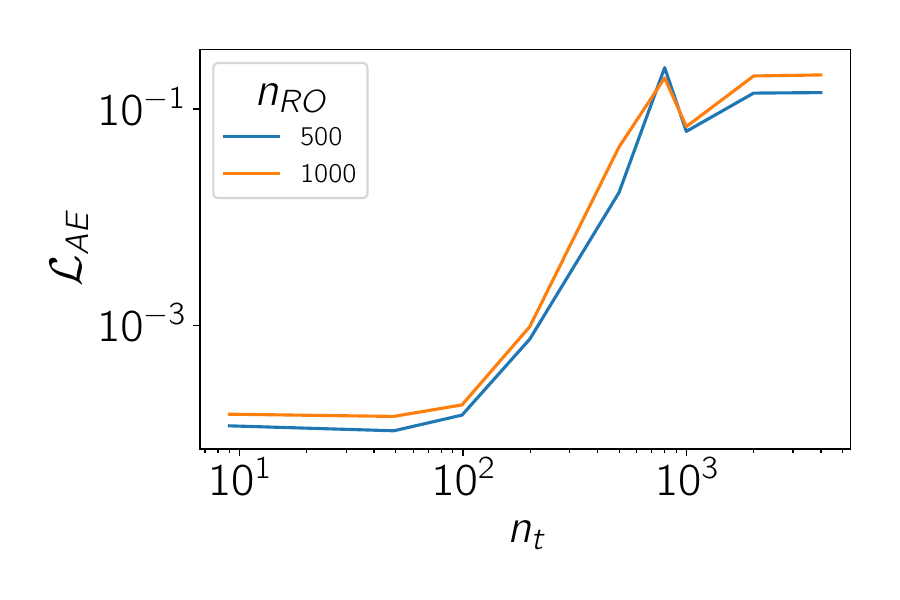}
  \caption{$\mathcal{L}_{AE}$ (Eq.~\ref{eq: Autoenc_loss}) as a function of $n_t$ using the coupled training approach with (left) and without (right) the $\mathcal{L}_{2}$ term in the combined loss (Eq.~\ref{eq: coupled training loss}).}
  \label{fig:CoupVsDecoup_L2Comp}
\end{figure}

Figure~\ref{fig:CoupVsDecoup_L2Comp} illustrates $\mathcal{L}_A$ for coupled autoencoders, comparing those trained with and without the $\mathcal{L}_{2}$ loss term, across different training trajectory lengths ($n_t$).

In both scenarios, it is noted that $\mathcal{L}_A$ increases with higher values of $n_t$. The relationship is attributed to the fact that increasing $n_t$ lengthens the trajectory predicted by the neural ODE, consequently enhancing the complexity of the $\mathcal{L}_1$ loss term in the coupled training loss (Eq. (\ref{eq: coupled_loss})), which could account for the observed trend in Fig.~\ref{fig:CoupVsDecoup_L2Comp}. Notably, throughout the range of $n_t$ values, coupled autoencoders trained with the $\mathcal{L}_{2}$ loss term exhibit lower instantaneous projection errors. Consequently, for subsequent comparative studies, a coupled autoencoder trained with the $\mathcal{L}_{2}$ loss term is utilized.

\begin{figure}
  \centering
  \includegraphics[width=0.49\textwidth]{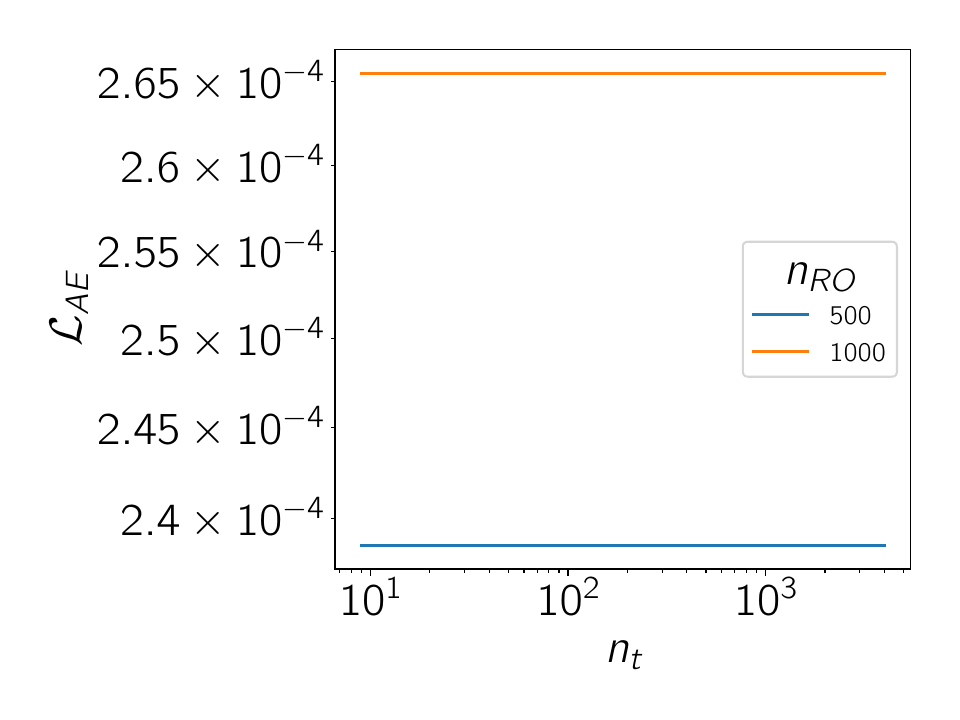}
  \includegraphics[width=0.49\textwidth]{Results/L2vL1_withL2.pdf}
  \caption{$\mathcal{L}_{\text{AE}}$ (Eq.~\ref{eq: Autoenc_loss}) as a function of $n_t$ using the (Left) de-coupled and (Right) coupled training approach.}
  \label{fig:CoupVsDecoup_projError}
\end{figure}

\textbf{Projection error}: Figure~\ref{fig:CoupVsDecoup_projError} presents a comparison of the autoencoder reconstruction error $\mathcal{L}_A$,  between autoencoders trained using decoupled and coupled approaches. This analysis considers various sample sizes (rollout-length) over a range of $n_t$ values. In the decoupled approach, a single autoencoder is trained, while different neural ODEs are trained with varying values of $n_t$. Consequently, the autoencoder projection errors remain independent of $n_t$ in this scenario. Conversely, in the coupled training approach discussed in the previous section, the instantaneous projection error increases with $n_t$. When examining the magnitudes of the projection error for autoencoders trained using decoupled and coupled approaches, it becomes apparent that the decoupled approach yields lower projection errors. This outcome aligns with expectations, considering that the coupled approach optimizes parameters for both the autoencoder and neural ODE simultaneously, posing a more challenging optimization problem in general.

\subsubsection{Training trajectory length}

The significance of another crucial hyperparameter, namely the training trajectory length $n_t$, was analyzed here. This parameter is chosen \textit{a priori} and has a substantial impact on the predictive accuracy of the system. Given that the focus of this work is on time-scale analysis in the latent space, the influence of both $n_t$ and certain hyperparameters related to network architecture of the autoencoder, on the latent time-scales, are isolated.

\begin{figure}
  \centering
  \includegraphics[width=0.49\textwidth]{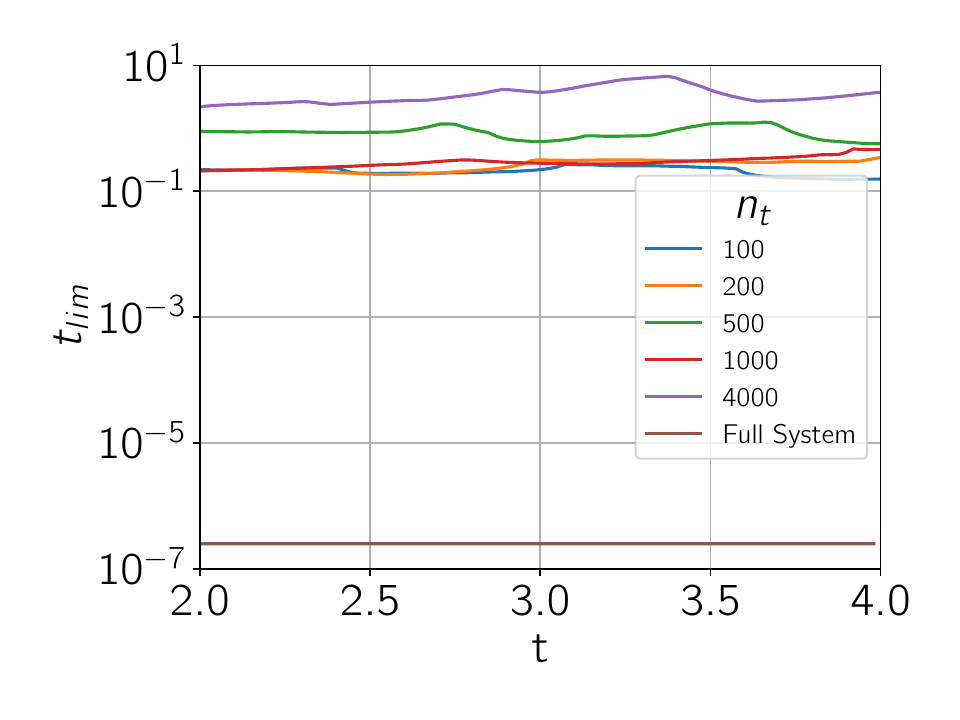}
  \includegraphics[width=0.49\textwidth]{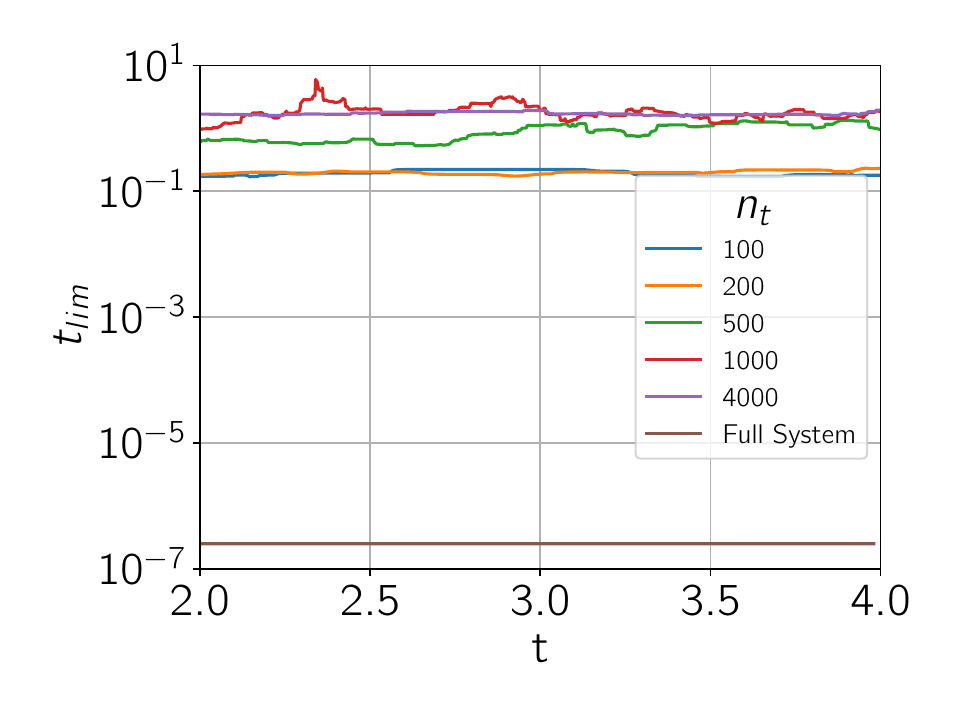}
  \caption{Limiting time-scale in the latent space ($t_{lim}$) as a function of time for neural ODEs with different training trajectory lengths ($n_{t}$) for the de-coupled (left) and coupled (right) training approaches. All models utilize latent dimensionality of 25 and four convolution layers.}
  \label{fig:tLim_nt}
\end{figure}

Figure~\ref{fig:tLim_nt} displays the limiting time-scale ($t_{lim}$) as a function of time for neural ODEs trained with different training trajectory lengths ($n_t$) in both the decoupled and coupled training approaches. The plot reveals that increasing $n_t$ results in an augmented limiting time-scale in the latent space, indicating smoother latent space trajectories that allow for larger time-steps. Specifically, transitioning from an $n_t$ value of 100 to 4000 leads to an increase in the limiting time-scale in the latent space by approximately two orders of magnitude.

When comparing the corresponding limiting time-scales for the coupled and decoupled training approaches, it becomes apparent that $t_{lim}$ exhibits similar magnitudes for a given $n_t$ value in both approaches. Consequently, the training methodology does not significantly impact the time-scale reduction in the latent space.

Summarizing the findings from previous sections, it is evident that among all the studied hyperparameters affecting time-scale reduction, the training trajectory length ($n_t$) is the sole parameter with a substantial impact on $t_{lim}$. Increasing $n_t$ results in smoother latent space trajectories, allowing for larger time-steps in the latent space. However, for predictions at unseen parameter instances, the relationship between increasing $n_t$ and accuracy is not straightforward, as discussed later. Consequently, there exists an optimal value for $n_t$ that strikes the best trade-off between predictive accuracy and time-scale reduction.

\begin{figure}
  \centering
  \includegraphics[width=0.49\textwidth]{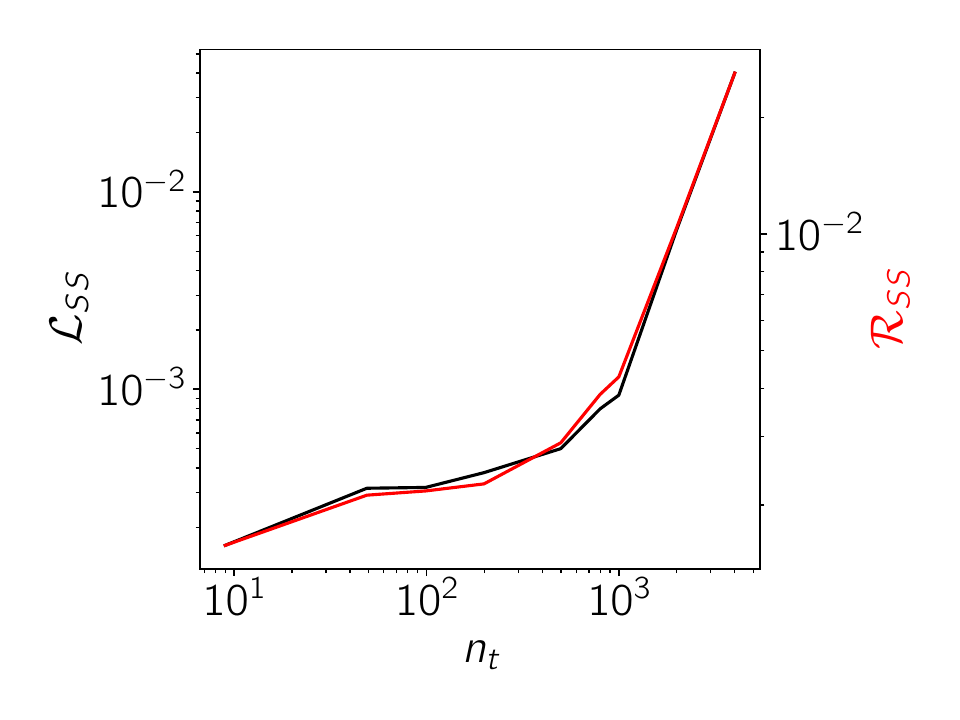}
  \includegraphics[width=0.49\textwidth]{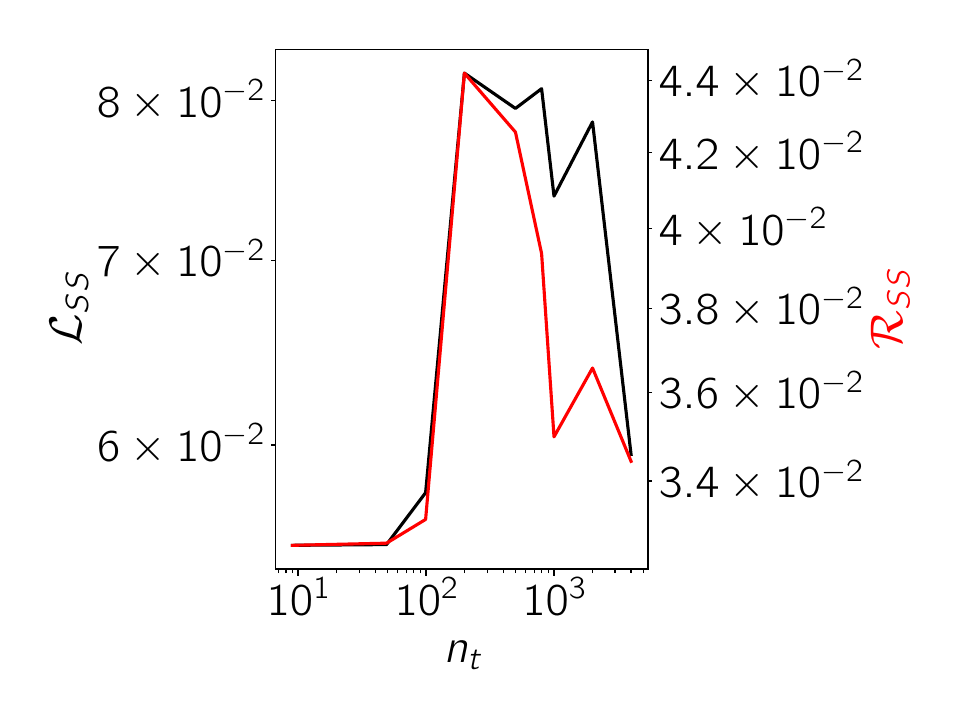}
  \caption{$\mathcal{L}_{\text{SS}}$ (Eq.~\ref{eq: SSR_error}) and $\mathcal{R}_{\text{SS}}$ (Eq.~\ref{eq: RSS Loss  }) as a function of $n_t$ using de-coupled (left) and coupled (right) training approaches. All models utilize latent dimensionality of 25 and four convolution layers.}
  \label{fig:CoupVsDecoup_singRollError}
\end{figure}

\textbf{Single step rollouts}: To assess the predictive accuracy of the combined autoencoder neural ODE framework while excluding the impact of error accumulation from extended rollouts, we calculate the Mean Squared Error (MSE) loss $\mathcal{L}_{\text{SSR}}$ (given by Eq. (\ref{eq: SSR_error})) and Relative Absolute Error (RAE) (Eq.~\ref{eq: RSS Loss })following a single-step prediction in time, considering an unseen initial condition. 

% The RAE is determined by the formula:

% \begin{equation} \label{eq: RSS Loss }
%     \mathcal{R}_{\text{SS}} = \Biggl \langle \frac{ \left| \psi(\mathbf{\widetilde{{w}}}(t+\Delta t)) - \mathbf{u}(t+\Delta t) \right| }{\left| \mathbf{u}(t+\Delta t) \right|} \Biggr \rangle,
% \end{equation}

providing a measure of the relative error in the predicted solution. Figure~\ref{fig:CoupVsDecoup_singRollError} illustrates the $\mathcal{L}_{\text{SSR}}$ and $\mathcal{R}_{\text{SSR}}$, comparing the decoupled and coupled training approaches across various $n_t$ values.

Across the entire range of $n_t$ values, the decoupled training approach exhibits lower MSE and RAE values. In the decoupled approach, both losses increase with $n_t$ as expected since the single-step prediction loss is optimized for $n_t = 1$. For the coupled approach, the losses initially rise with $n_t$ and then decrease for $n_t$ values exceeding 200.

\textbf{Rollouts}: To compare the predictive accuracy of different models over longer rollout trajectory lengths, the Mean Squared Error $\mathcal{L}_{\text{RO}}$ (given by Eq. (\ref{eq: RO_error}))and Relative Absolute Error $\mathcal{R}_{\text{RO}}$ (Eq.~\ref{eq: RSS Loss }) are used.

\begin{figure}
  \centering
  \includegraphics[width=0.49\textwidth]{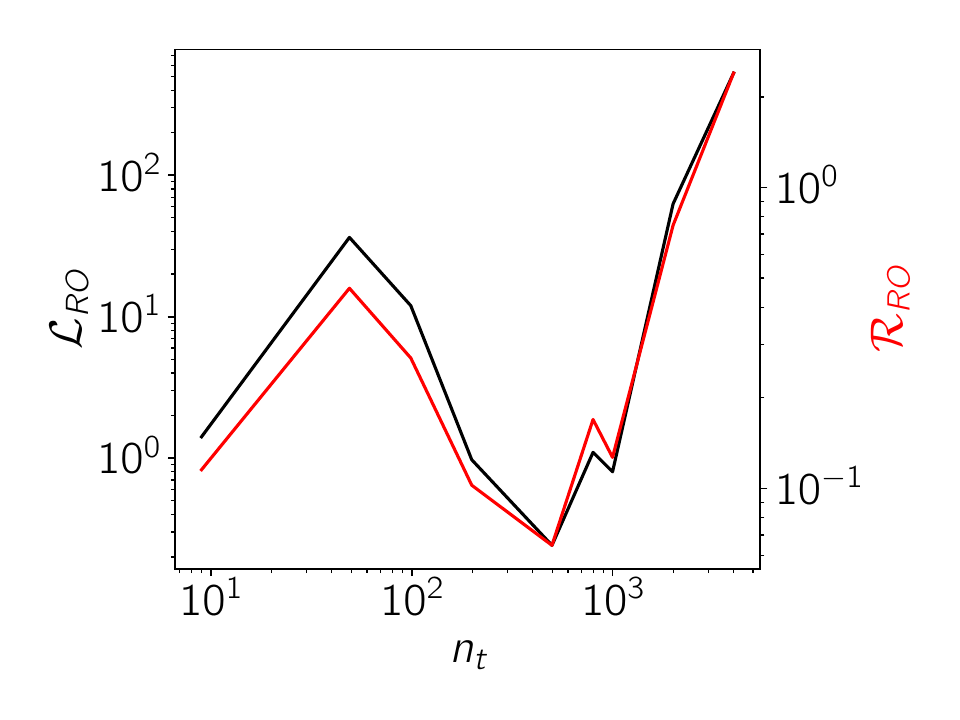}
  \includegraphics[width=0.49\textwidth]{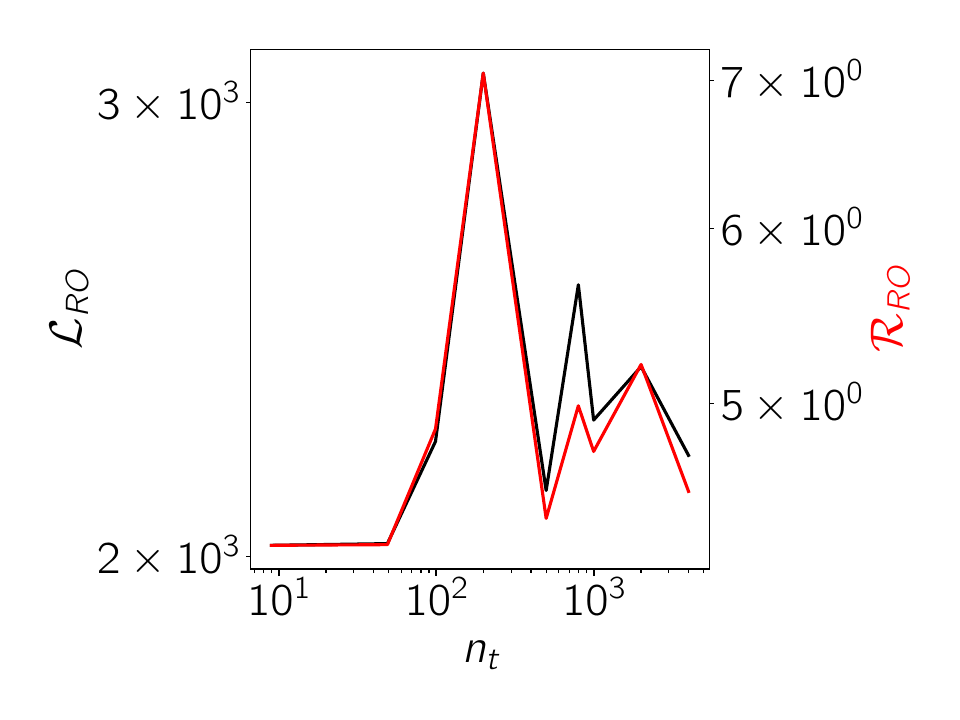}
  \caption{$\mathcal{L}_{\text{RO}}$ (Eq.~\ref{eq: RO_error}) and $\mathcal{R}_{\text{RO}}$ (Eq.~\ref{eq: RRO Loss }) for a rollout trajectory length $n_{\text{RO}} = 500$ as a function of $n_t$ using de-coupled (left) and coupled (right) training approaches. Models utilize latent dimensionality of 25 and four convolution layers.}
  \label{fig:CoupVsDecoup_RollError}
\end{figure}

% given by,
% \begin{equation} \label{eq: RRO Loss }
%     \mathcal{R}_{\text{RO}} = \Biggl \langle \sum^{n_{RO}}_{j=1} \frac{ \left| \psi(\mathbf{\widetilde{{w}}}(t + \Delta t)) - \mathbf{u}(t + j \Delta t) \right| }{\left|  \mathbf{u}(t + j \Delta t) \right|} \Biggr \rangle,
% \end{equation}
Figure~\ref{fig:CoupVsDecoup_RollError} depicts a comparison of $\mathcal{L}_{\text{RO}}$ and $\mathcal{R}_{\text{RO}}$ for a rollout trajectory length $n_{\text{RO}}=500$ timesteps in both the coupled and decoupled approaches across a range of $n_t$ values. Notably, for the extended rollout trajectory, the decoupled approach demonstrates significantly lower error levels, nearly two orders of magnitude less, in comparison to the coupled approach.

In the case of the coupled approach, there is no discernible dependence of the rollout errors on $n_t$, except that lower $n_t$ values correspond to lower rollout errors. This observation can be explained by noting that the predictive accuracy of the autoencoder and neural ODEs, trained using the coupled approach, appears to be diminishing as more samples are utilized for training (higher values of $n_t$).

Conversely, for the decoupled approach, the rollout losses initially increase with $n_t$ and then reach their minimum at an optimal $n_t$ of 500. This implies that the combined decoupled neural ODE is most adept at representing the underlying physics of the system at this specific value of $n_t$.  Although this may seem expected, the identification of this 'optimal' $n_t$ holds substantial significance for achieving the best predictive accuracy. Subsequent sections explore a method to determine this optimal $n_t$ \textit{a priori}, providing insights into achieving the highest predictive accuracy.

\subsubsection{Largest time-scales}

In addition to investigating the limiting time-scales in the latent space, we also explore the largest time scales, represented by the inverse of the smallest eigenvalues of the right-hand side Jacobian and denoted as $t_{max}$ (defined in Section~\ref{Sec: Dyn sys time-scales}). The largest time-scales of a dynamical system typically contain the majority of the system's energy, and ideally, a neural ODE that captures these large time-scales should yield better predictive accuracy.

\begin{figure}
  \centering
  \includegraphics[width=0.49\textwidth]{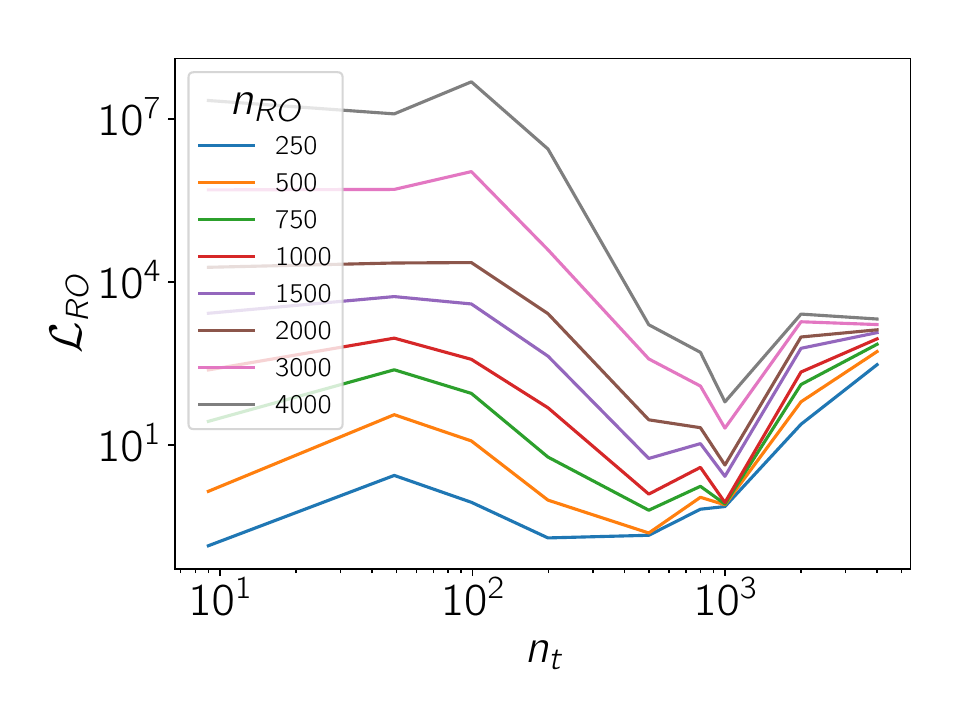}
  \includegraphics[width=0.49\textwidth]{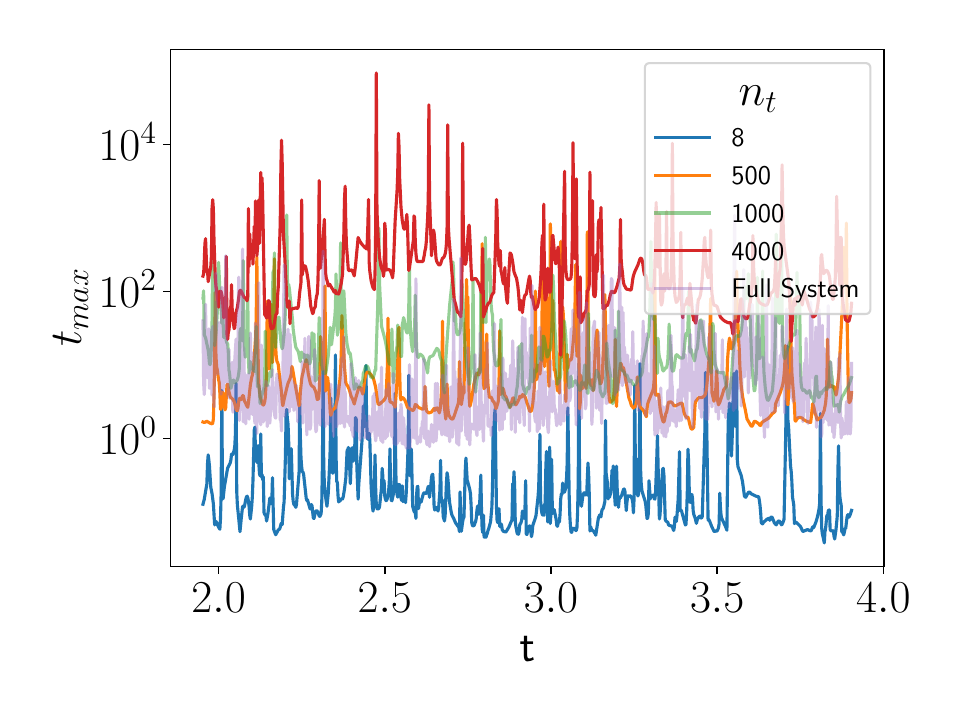}
  \caption{\textbf{(Left)} $\mathcal{L}_{\text{RO}}$ (Eq.~\ref{eq: RO_error}) as a function of $n_t$ for different rollout trajectory lenghts $n_{\text{RO}}$ using the de-coupled training strategy. \textbf{(Right)} Largest time-scale in the latent space ($t_{max}$) (Eq.~\ref{eq: nODE tlim tmax}) as a function of time for neural ODE with different training trajectory lengths ($n_{t}$). Models utilize latent dimensionality of 25 and four convolution layers. }
  \label{fig:tmax}
\end{figure}

Figure~\ref{fig:tmax} (left) shows $\mathcal{L}_{\text{RO}}$ as a function of $n_t$ for different rollout trajectory lengths $n_{\text{RO}}$. It can be seen that for $n_{\text{RO}}=500$ and lower, the lowest $\mathcal{L}_{\text{RO}}$ is seen near $n_t=500$. As $n_{\text{RO}}$ is increased beyond 500, the minimum shifts to $n_t = 1000$ and this trend remains consistent for all higher $n_{\text{RO}}$ values, suggesting that neural ODEs with $n_t=500$ and $n_t=1000$ have the highest predictive accuracy for rollout predictions.

% the neural ODE with $n_t=7$ and $n_t=3000$ has $t_{max}$ values lower and higher than that of the full system, respectively. In contrast, 

To understand why this is the case, Fig.~\ref{fig:tmax} (right) illustrates $t_{max}$ as a function of time for neural ODEs with different training trajectory lengths. The plot reveals that the neural ODEs trained with $n_t=500$ and $n_t=1000$ have $t_{max}$ values closely aligning with those of the full system, in contrast to higher (4000) and lower (8) $n_t$ values. As observed in Fig.~\ref{fig:CoupVsDecoup_RollError} and \ref{fig:tmax}, since $n_t=500$ and $n_t=1000$ yield the lowest rollout errors for unseen parameter instances, this observation suggests that neural ODEs with the best predictive accuracy have $t_{max}$ values closely aligned with those of the full system, and therefore effectively capture the slowest evolving physics.

While the precise determination of an optimal value for $n_t$ remains unclear, the results shown here suggest that examining neural ODE behavior on the basis of $t_{max}$ can offer insights into the range of $n_t$ values that would likely result in the best predictive accuracy.

\subsection{Extension to Channel Detonations} \label{Sec: 1D Detonations}

In this section, to further demonstrate the generality of the timescale relationships described in the context of the KS equations in the previous sections, additional studies of the neural ODE strategy are performed on the compressible reacting Navier-Stokes equations. More specifically, surrogate models are constructed for unsteady gaseous detonations in a one-dimensional channel configuration, constituting a more realistic and highly advection-dominated demonstration case that incorporates multiple transported variables. In broad terms, an unsteady detonation can be interpreted as a propagating shockwave coupled with a chemical reaction zone \cite{shepherd}. Accelerated simulations of detonation-containing flows are of significant interest to the high-speed propulsion community, where emerging concepts reliant on detonation-based combustion offer pathways for higher efficiency and robust designs~\cite{shepherd,venkat_arfm_2023}. In this context, similar to the KS demonstration studies in the previous sections, numerical solutions of the compressible NS equations with detailed chemical kinetics are used to generate ground-truth trajectory data describing the propagation of self-sustained detonation waves, from which the combined autoencoder-neural ODE models are trained. The objective is to not only showcase the capability of neural-ODE based approaches in modeling detonation dynamics, but also to illustrate consistency in neural ODE timescale trends (namely, the relationship between $n_t$ and the latent timescales) across fundamentally different PDEs. 

\begin{figure}
    \centering
    \includegraphics[width=\columnwidth]{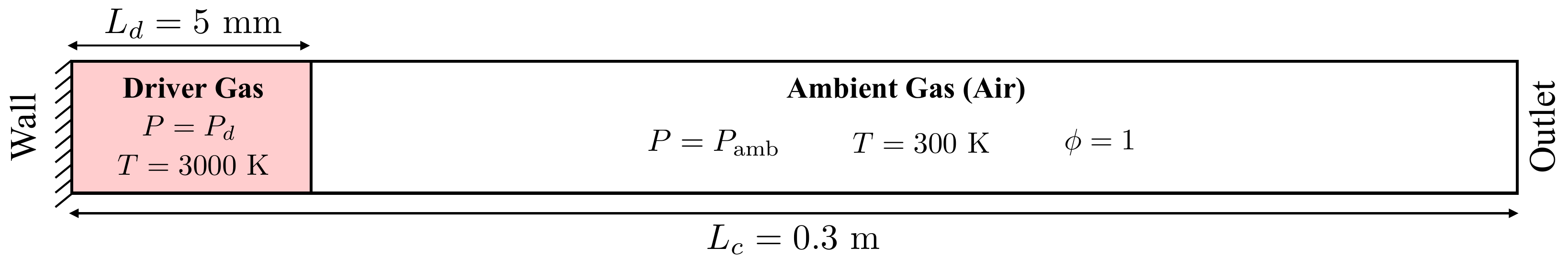}
    \caption{Schematic of channel detonation configuration and initial condition, where a high-energy driver gas is used to initialize a self-sustained detonation wave that propagates left-to-right through the channel. Ambient and driver pressures for the trajectories considered are provided in Table~\ref{table:detonation_cases}. Driver gas compositions (not shown in schematic) come from Chapman-Jouguet conditions obtained from the Shock and Detonation toolbox \cite{sdtoolbox}. Initial fluid velocity is zero throughout entire domain.}
    \label{fig:detonation_config}
\end{figure}

The detonation configuration and initiation strategy is shown in Fig.~\ref{fig:detonation_config}. More specifically, unsteady Hydrogen-Air detonations are initialized in the manner of Ref.~\cite{yungster2004pulsating}, where a driver gas at elevated pressure and temperature near the left wall is used to establish a self-sustained Chapman-Jouguet detonation wave that propagates through the channel. {To generate ground-truth detonation data to train the data-based models, the compressible reacting NS equations are solved using a flow solver developed at the University of Michigan based on the AMReX framework \cite{shivank_paper,amrex}}. The solver is a block-structured adaptive mesh refinement (AMR) extension of the extensively verified UMReactingFlow \cite{umreactingflow} (note that although an AMR-based solver is employed here, grid refinement is not used in this study). A globally second-order finite-volume strategy is utilized, where advection terms are treated with slope-limited Harten-Lax-van Leer-Contact approximate Riemann solver \cite{batten_1997} and diffusion terms are treated using standard central schemes. Detailed chemistry kinetics routines (species production rate and transport coefficient evaluations) are handled by Cantera \cite{cantera}. In this work, hydrogen-air chemistry is modeled using the 9 species, 21 reaction detailed mechanism of Mueller et al. (1999) \cite{hon_mueller}. Global time integration is handled using a Strang splitting strategy; chemical time integration is performed using an adaptive explicit method, and the advection-diffusion temporal advance uses a stability-preserving second-order Runge-Kutta method. The reader is directed to Ref.~\cite{umreactingflow} for additional detail on the solver numerics and discretization approach. 

Detonation dynamics are parameterized by both ambient gas and driver gas properties. In particular, the ambient gas is known to control detonation wave speeds, peak pressures, and chemical timescales in the wave structure (i.e., higher ambient gas pressures result in smaller chemical timescales and more chemically stiff wave structures), and the driver gas has additional, albeit more minor, effects on detonation coupling and observed wave speeds. As a result, parameters of the the trajectories used to train the models are described by two distinct driver pressure ratios and ambient pressures, and are provided in Table~\ref{table:detonation_cases}. For both the decoupled autoencoder and neural ODE, only half of the total time snapshots from each trajectory are utilized for training purposes. In each snapshot, each spatial discretization point stores fluid density, pressure, temperature, velocity, and all species mass fractions, resulting in an input snapshot channel depth of 13 (in contrast to the unity depth used in the KS equation). 

\begin{table}[h!]
\centering
\begin{tabular}{||c c c c c c c||} 
 \hline
 Trajectory & $P_{\text{amb}}$ & $T_{\text{amb}}$ & $P_d/P_{\text{amb}}$ & $T_d/T_{\text{amb}}$ & Training Snapshots & Testing Snapshots \\ [0.5ex] 
 \hline\hline
 1 & 0.5 atm & 300 K & 40 & 10 & 7500 & 7500 \\ 
 \hline
 2 & 1 atm & 300 K & 20 & 10 & 7500 & 7500 \\ 
 \hline
\end{tabular}
\caption{Detonation dataset description.}
\label{table:detonation_cases}
\end{table}

\begin{figure}
  \centering
  \includegraphics[width=0.60\textwidth, height=8cm]{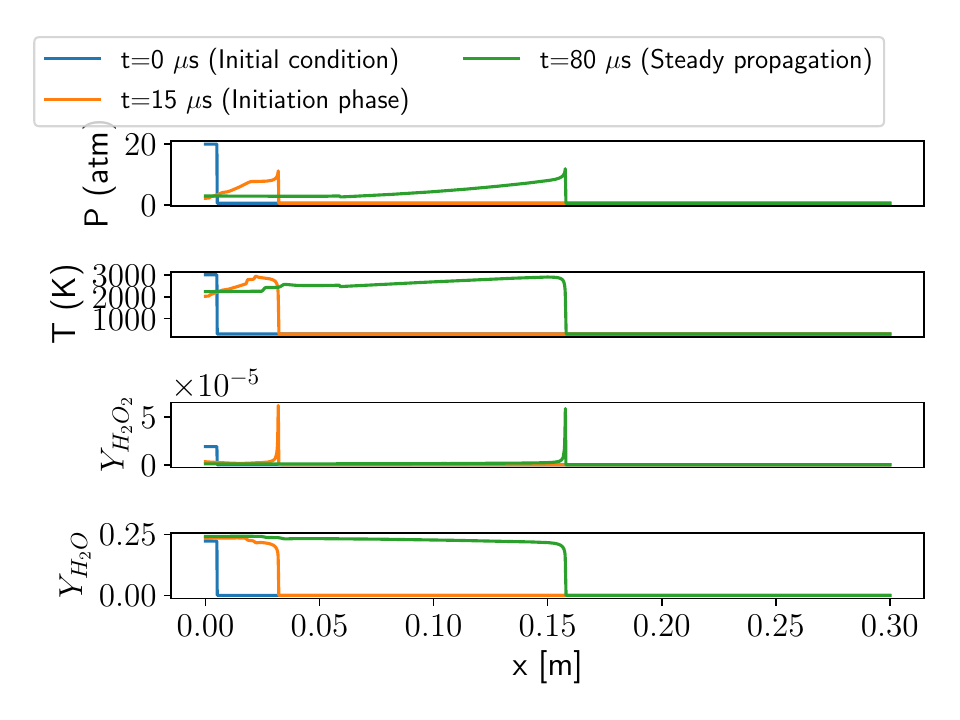}
  \includegraphics[width=0.39\textwidth]{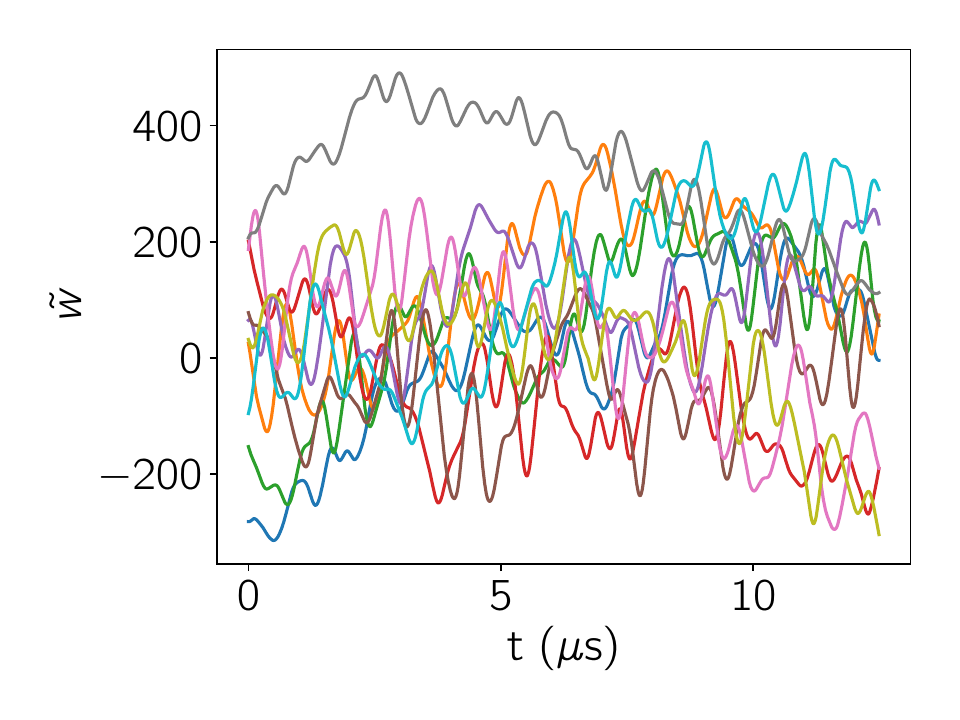}
  \caption{(Left) Pressure, temperature and species mass fractions fields for Trajectory 1 at different time instances showing the initial condition, initiation phase and self-sustained propagation phase. (Right) Latent space trajectories obtained during self-sustained propagation phase (with $n_t=250$, $N_w=10$ and four convolutional layers in the autoencoder).}
  \label{fig: waveProf_latTraj}
\end{figure}

\begin{figure}
  \centering
  \includegraphics[width=0.49\textwidth]{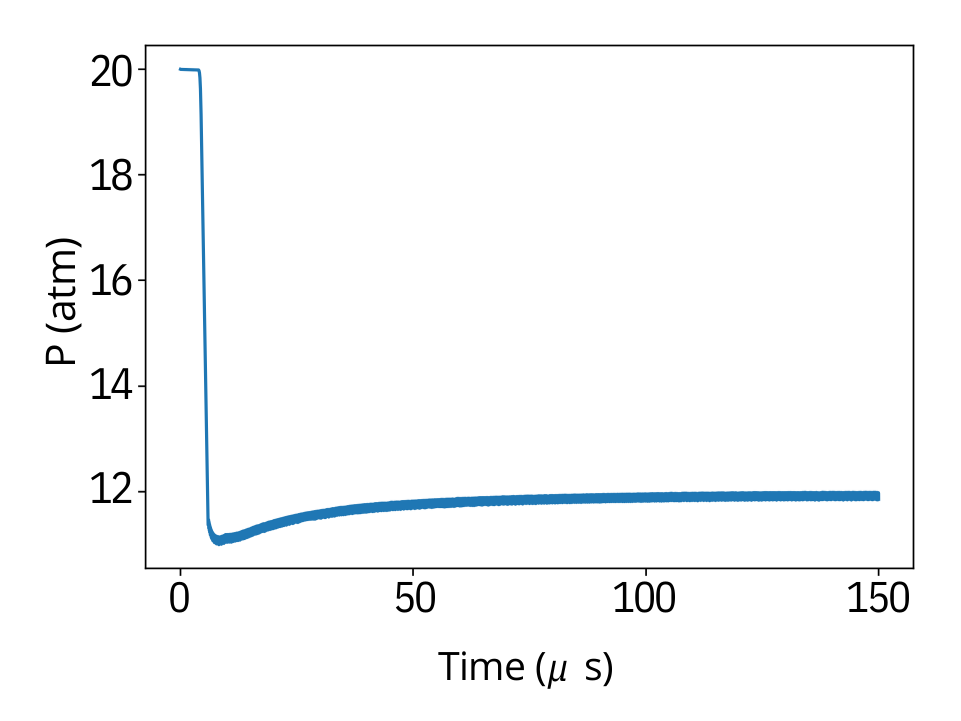}
  \includegraphics[width=0.485\textwidth]{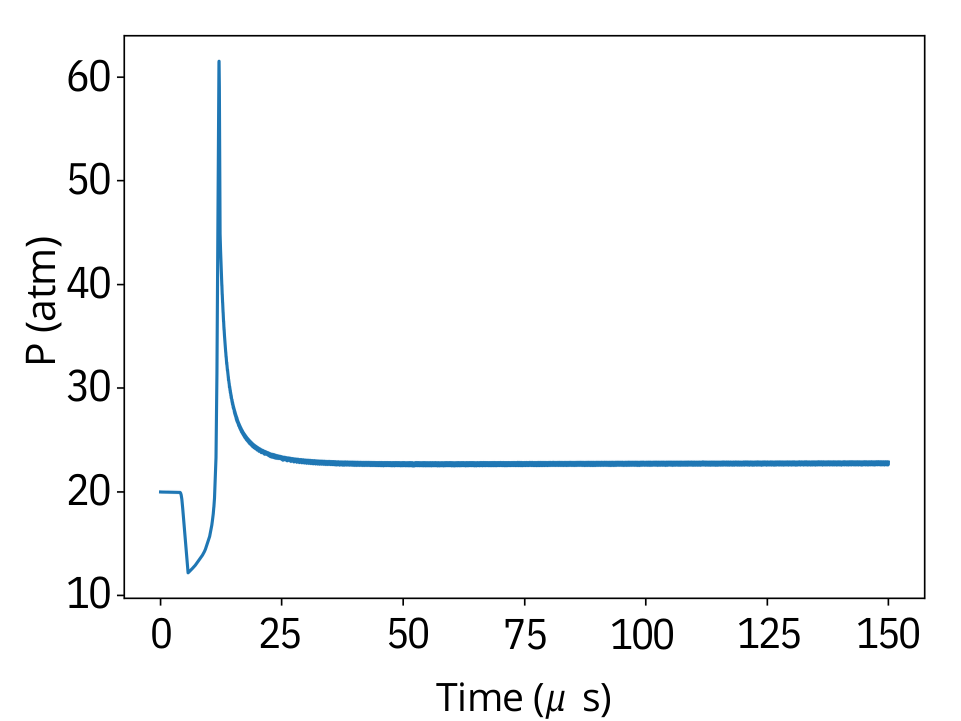}
  \caption{Peak pressure traces for Trajectory 1 and 2 (left and right, respectively). See Table~\ref{table:detonation_cases} for trajectory descriptions.}
  \label{fig: presProbe}
\end{figure}

For all ground-truth simulation trajectories, the detonation channel length was set to $0.3$~m, the grid resolution was fixed to 50~$\mu$m. This resolution was sufficient enough to allow for proper formation of a self-sustained detonation, allowing all complex reacting flow dynamics to be captured. The simulation time-step was fixed to $5$~ns, and snapshots were written at intervals of $\Delta t = 10^{-8}$ for neural ODE training and inference. Examples of detonation wave evolution, along with the corresponding smooth latent space trajectories are shown in Figure~\ref{fig: waveProf_latTraj}. Figure~\ref{fig: presProbe} shows the peak pressure traces for both the considered trajectories. 

\subsubsection{Extrapolation in time}

\begin{figure}
  \centering
  \includegraphics[width=0.49\textwidth, height=7cm]{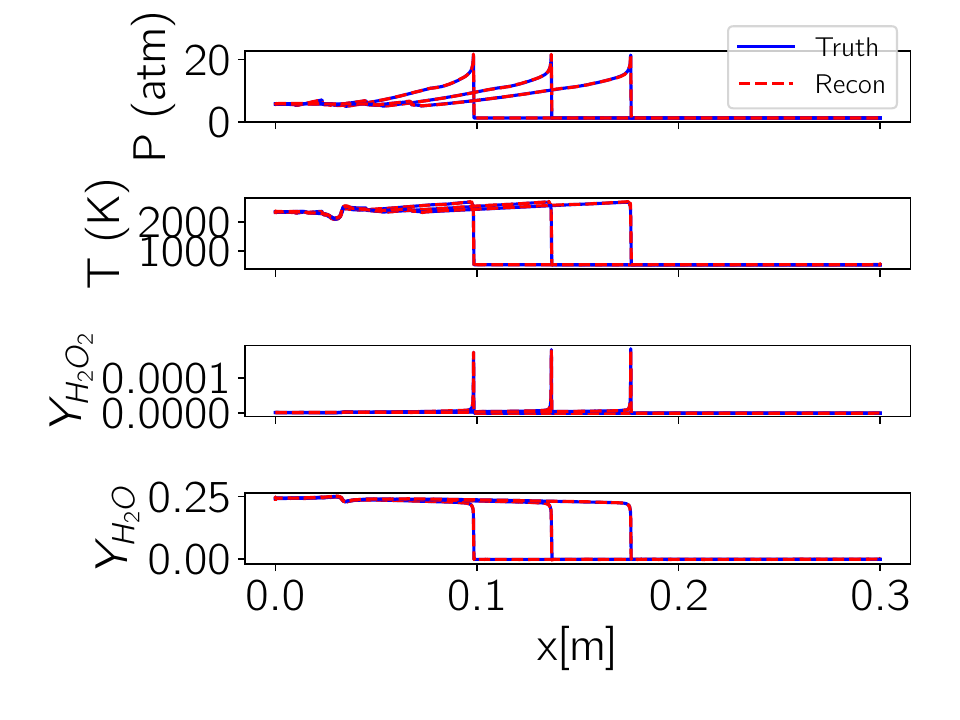}
  \includegraphics[width=0.49\textwidth, height=7cm]{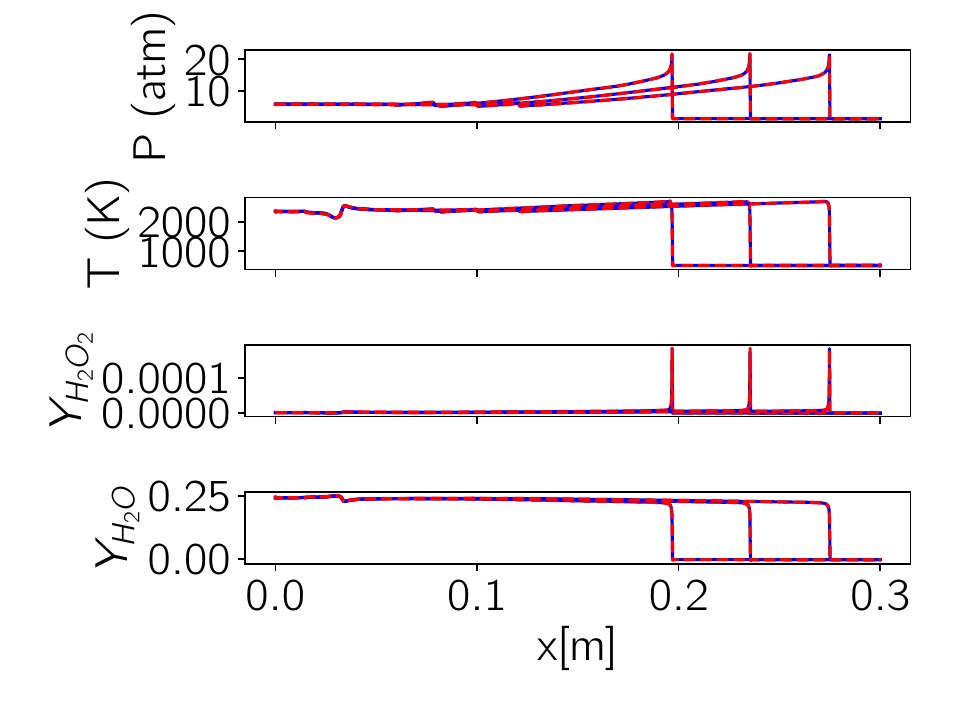}
  \caption{Comparison of the ground truth and predicted fields for Trajectory 1 within the training set (left) and outside the training set (right). Results shown for $n_t=250$,  $N_w=10$, and four convolutional layers, with extrapolation time of 37.5 $\mu s$) (7500 time-steps)}
  \label{fig:Det_extInT_1}
\end{figure}

\begin{figure}
  \centering
  \includegraphics[width=0.49\textwidth, height=7cm]{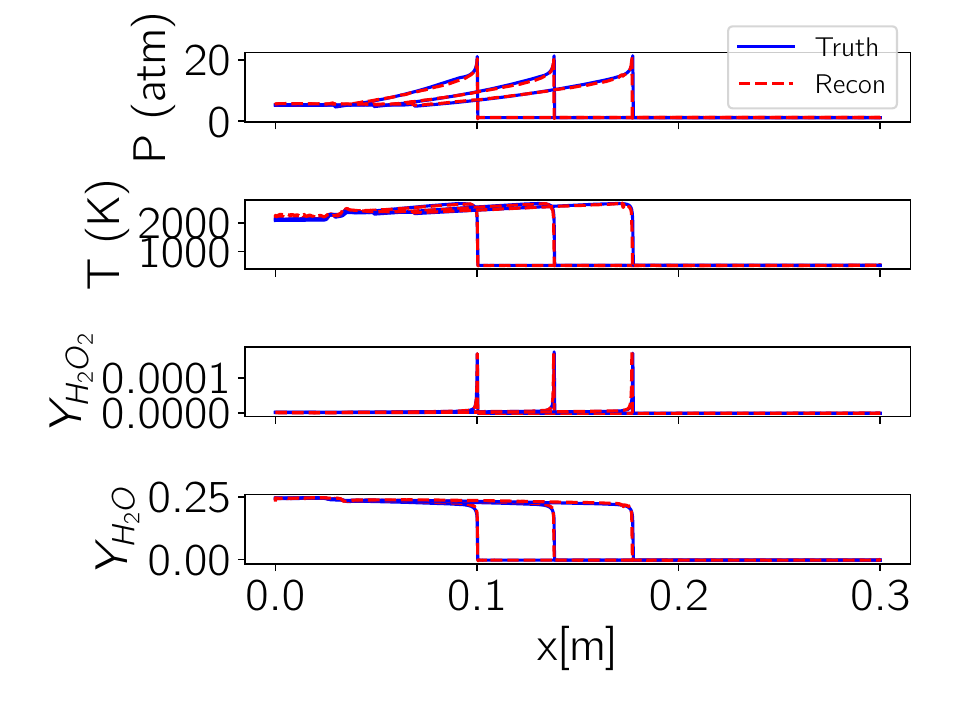}
  \includegraphics[width=0.49\textwidth, height=7cm]{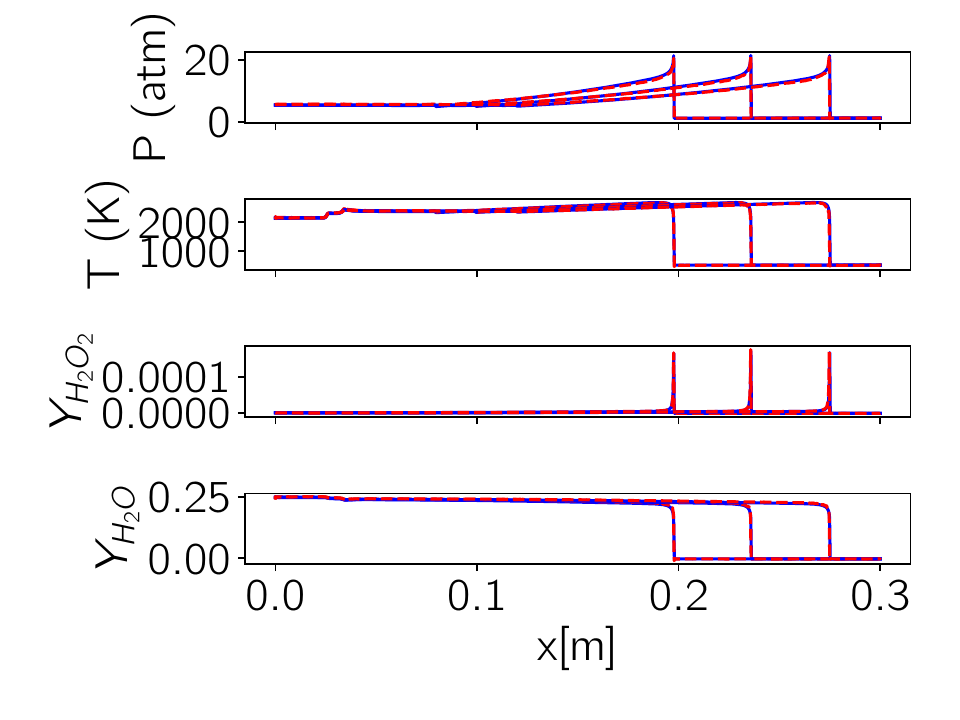}
  \caption{Comparison of the ground truth and predicted fields for Trajectory 2 within the training set (left) and outside the training set (right). Results shown for $n_t=125$,  $N_w=10$, and four convolutional layers, with extrapolation time of 37.5 $\mu s$) (7500 time-steps))}
  \label{fig:Det_extInT_2}
\end{figure}

The predictive accuracy of the trained autoencoder neural ODE framework trained using the decoupled approach and with $n_t=250$, $n_{\hat{u}}=10$ and four convolutional layers in the autoencoder (see Figure~\ref{fig: nODE_autoenc_schem}), is evaluated by extrapolating in time beyond the training dataset, extending until the shock and flame exit the domain. In Fig.~\ref{fig:Det_extInT_1} and Fig.~\ref{fig:Det_extInT_2}, the predicted and ground truth normalized pressure, temperature, and mass fractions for $H_20_2$ and $H_20$ profiles are presented at time instances within and outside the training set. The results indicate that the neural ODE autoencoder framework closely aligns with the ground truth data within the training dataset and effectively extrapolates in time for both driver pressure ratios and ambient pressures.

\subsubsection{Effect of network hyperparameters}

\begin{figure}
  \centering
  \includegraphics[width=0.49\textwidth]{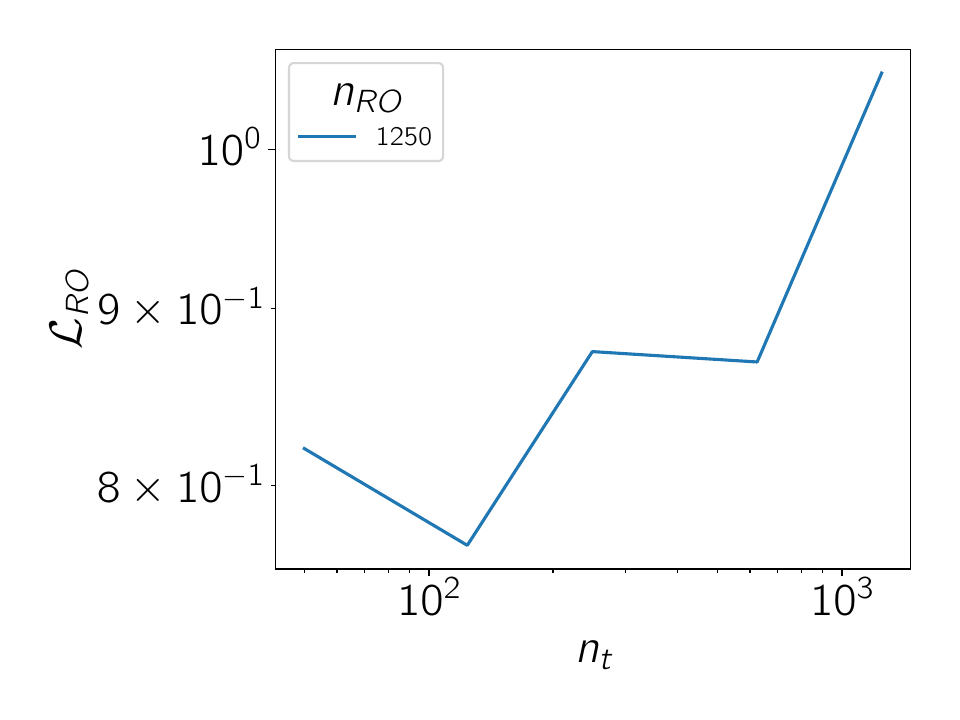}
  \caption{$\mathcal{L}_{\text{RO}}$ (Eq.~\ref{eq: RO_error}) for a rollout trajectory length $n_{\text{RO}} = 1250$, as a function of $n_t$ ($N_w=10$ and four convolutional layers).}
  \label{fig:1dDet_LossvsNt}
\end{figure}

Figure~\ref{fig:1dDet_LossvsNt} displays the Mean Squared Error (MSE) of predicted trajectories beyond the training set, calculated over 1250 samples, as a function of $n_t$. The graph reveals that the error is minimized at an optimal value of $n_t$, similar to what is observed in Figure~\ref{fig:CoupVsDecoup_RollError} for the KS equations.

\begin{figure}
  \centering
  \includegraphics[width=0.49\textwidth]{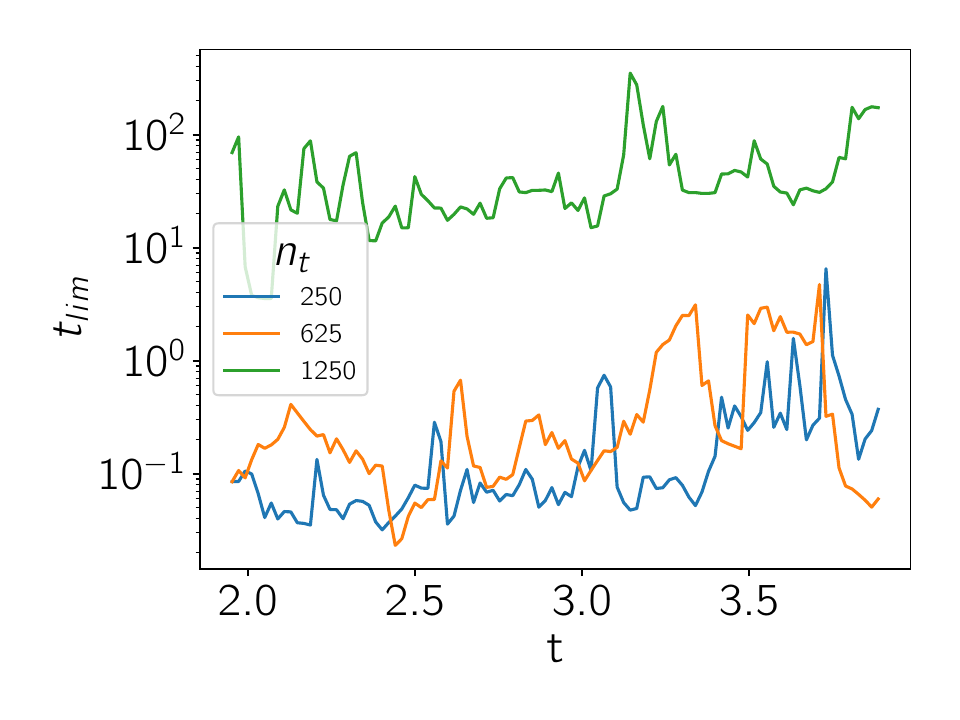}
  \caption{Limiting time-scale in the latent space ($t_{lim}$) (Eq.~\ref{eq: nODE tlim tmax}) as a function of time for neural ODEs with different training trajectory lengths $n_{t}$ ($N_w=10$ and four convolutional layers)}.
  \label{fig:1dDet_timeScaleComp}
\end{figure}

Figure~\ref{fig:1dDet_timeScaleComp} illustrates the limiting time-scale in the latent space, denoted as $t_{lim}$, over time for neural ODEs trained with different values of $n_t$. The graph indicates that increasing $n_t$ leads to higher values for $t_{lim}$, resulting in smoother latent space trajectories. This trend aligns with the observations for the KS equations shown in Figure.~\ref{fig:tLim_nt}. 
Comparing the latent space's limiting ($t_{lim}(t)$) and largest ($t_{max}(t)$) time-scales to their physical space counterparts, as in the KS equation, poses a challenge because the code generating the data prevents differentiation through the full-order dynamical system, essential for calculating Jacobians ($\frac{\partial F}{\partial u}$) required to determine the full-system time-scales.

\subsection{Extension to 2D Atmospheric flow} \label{Sec: 2D Atmospheric Flow}

In this section, additional demonstrations on an atmospheric flow dataset are performed to augment the results from the KS equation and detonation test cases, and to further verify trends with respect to the training trajectory length on the neural ODE strategy. Similar to Sec.~\ref{Sec: 1D Detonations}, the details of the dataset are first provided, followed by an assessment of model performance and $n_t$ effects. Additionally, this section also adds to the preexisting analysis above, in that the effect of simulation time-step variations on neural ODE performance is also assessed. 

The 2D atmospheric fields are produced using a simplified atmospheric general circulation model (AGCM) called SPEEDY. SPEEDY is a spectral transform AGCM that was developed to produce rapid climate simulations, using simplified, but modern physical parameterization schemes \cite{Molteni2003,Kucharski2006}. The grid of SPEEDY consists of a horizontal spatial resolution of  3.75$^\circ \times$3.75$^\circ$ with variables defined at eight vertical levels. The 3D varying state variables of the model are the two components of the horizontal wind vector, temperature, and specific humidity (moisture), while the single two-dimensionally varying state variable is the natural logarithm of surface pressure. For this dataset, we subsample the 3D variables to obtain 4 key variables of the model: near surface temperature, specific humidity, and upper level winds. 

This atmospheric dataset constitutes a highly complex and nonlinear advection-dominant benchmark that is fundamentally different from both KSE and detonation cases described above. It therefore serves as an additional dynamical system in this work to more comprehensively assess neural ODE timescale trends. For additional details on the physical phenomena contained in the dataset, the reader is directed to Refs.~\cite{Molteni2003,Kucharski2006}. 

%{\color{blue}description of data here}

\subsubsection{Extrapolation in time}

\begin{figure}
  \centering
  \includegraphics[width=0.49\textwidth]{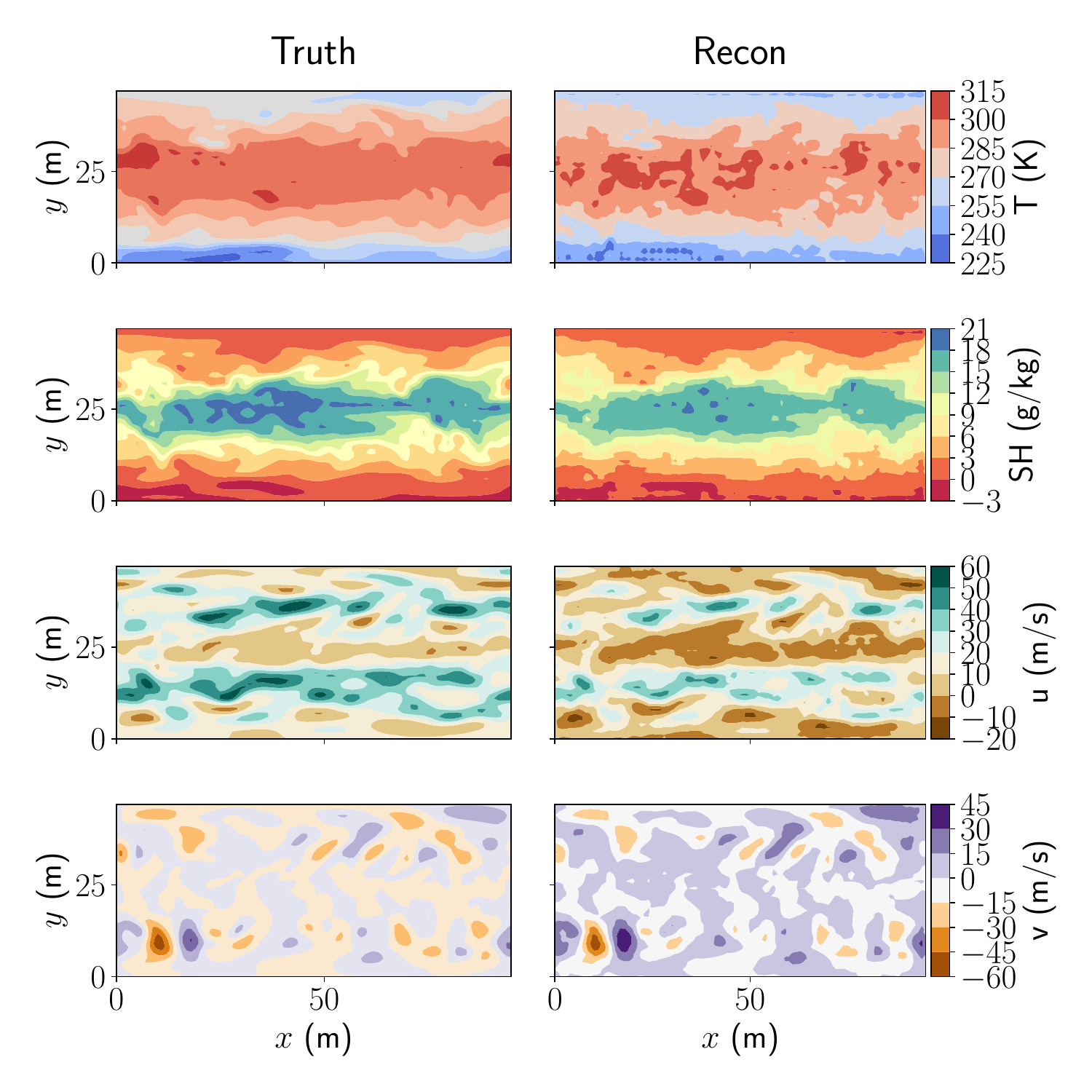}
  \includegraphics[width=0.49\textwidth]{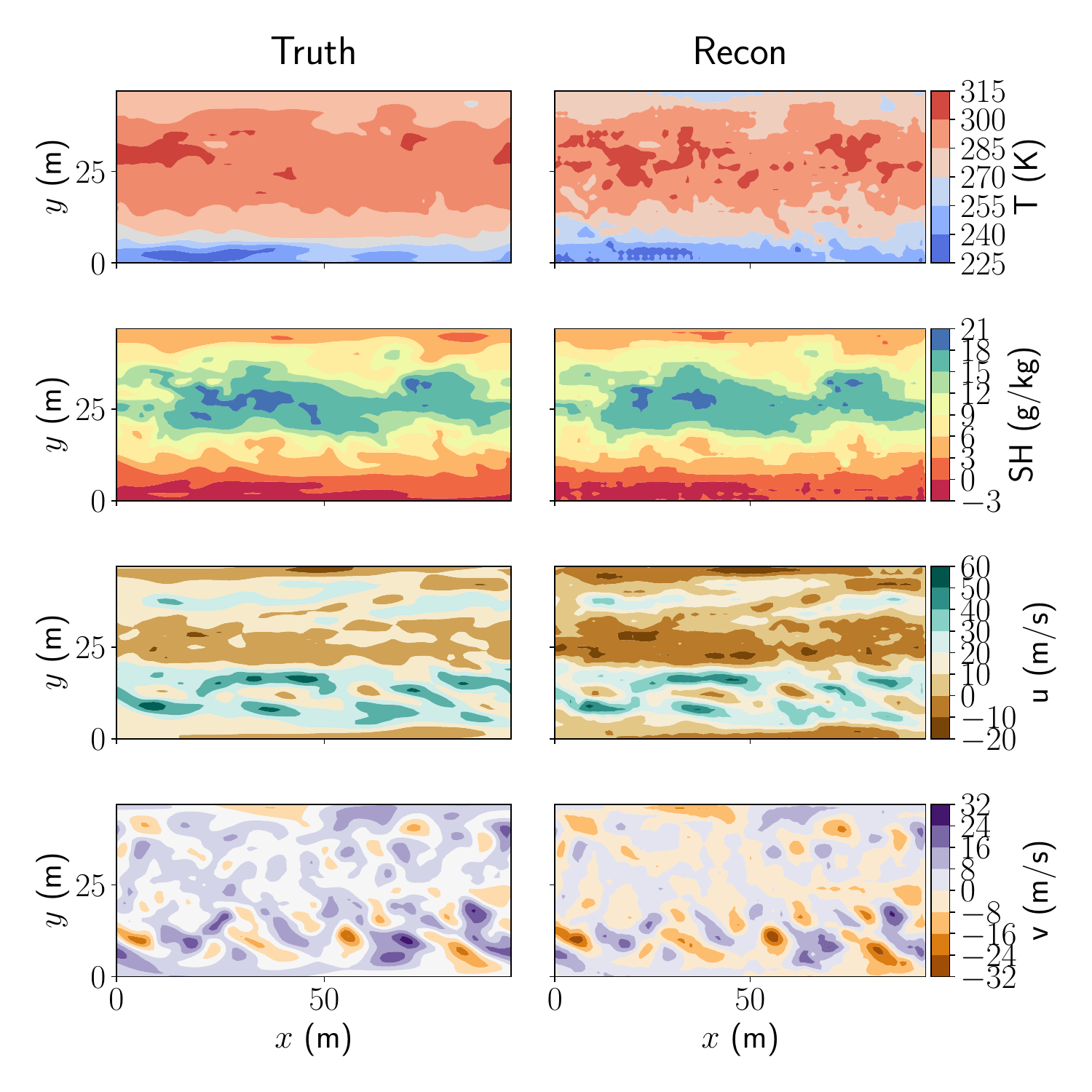}
  \caption{Comparison of the ground truth and predicted fields for atmospheric dataset the  within the training set (left) and outside the training set (right). Results shown for $n_t=20$,  $N_w=200$, and four convolutional layers, with extrapolation time of 2500 time-steps)}
  \label{fig:2D_extI}
\end{figure}

The predictive performance of the trained autoencoder neural ODE framework, developed using the decoupled approach and configured with \( n_t=20 \), \( n_{\hat{u}}=200 \), and four convolutional layers in the autoencoder (see Figure~\ref{fig: nODE_autoenc_schem}), is assessed by extrapolating beyond the training dataset. Figure~\ref{fig:2D_extI} presents the predicted and ground truth fields for temperature, specific humidity (SH), and the $u$ and $v$ velocity components at various time instances, both within and beyond the training period. The results suggest that the neural ODE autoencoder framework captures the system dynamics within the training domain and shows an ability to predict system behavior at unseen time instances. It should be noted that while many of the large scale features are preserved at the shown extrapolation time, there are indeed unphysical fine-scale artifacts in the predicted solution, pointing to intrinsic instabilities for this problem. Stability challenges for this dataset are generally expected using CNN-based encodings, due to the high level of nonlinearity and the fact that no additional stability enhancements are employed here. Despite this, it is emphasized that the purpose of this additional demonstration is to augment the KS equation and detonation analysis through inspection/understanding of the role of $n_t$, which is summarized below. 

\subsubsection{Effect of network hyperparameters}

\begin{figure}
  \centering
  \includegraphics[width=0.49\textwidth]{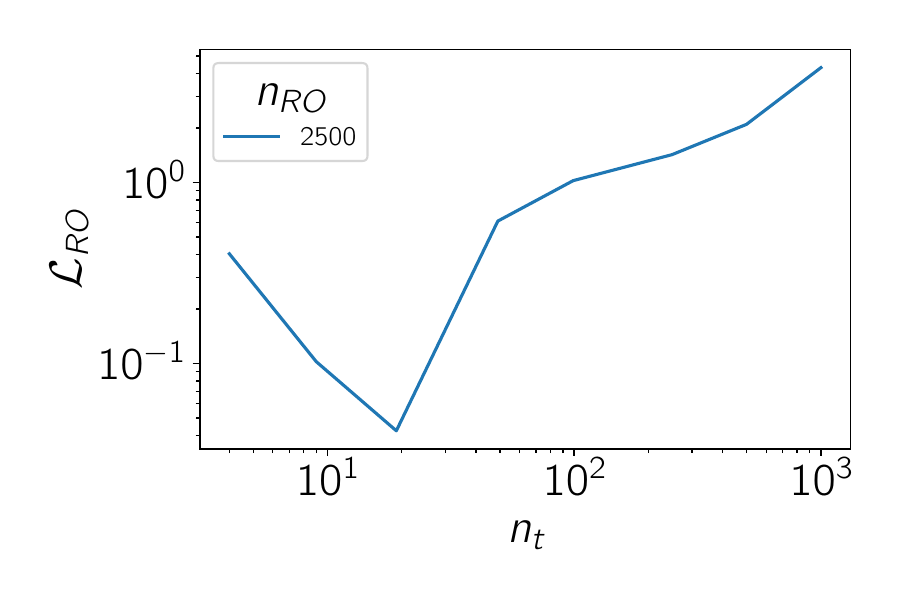}
  \caption{$\mathcal{L}_{\text{RO}}$ (Eq.~\ref{eq: RO_error}) for a rollout trajectory length $n_{\text{RO}} = 1000$, as a function of $n_t$ ($N_w=100$ and four convolutional layers).}
  \label{fig:2DAtmos_LossVsNt}
\end{figure}

Figure~\ref{fig:2DAtmos_LossVsNt} illustrates the Mean Squared Error (MSE) of predicted trajectories beyond the training set, evaluated over 1000 samples, as a function of $n_t$. The plot highlights that the error reaches its minimum at an optimal $n_t$, consistent with trends observed in Figure~\ref{fig:CoupVsDecoup_RollError} for the KS equations and Figure~\ref{fig:1dDet_LossvsNt} for the 1D channel detonation case.

\begin{figure}
  \centering
  \includegraphics[width=0.49\textwidth]{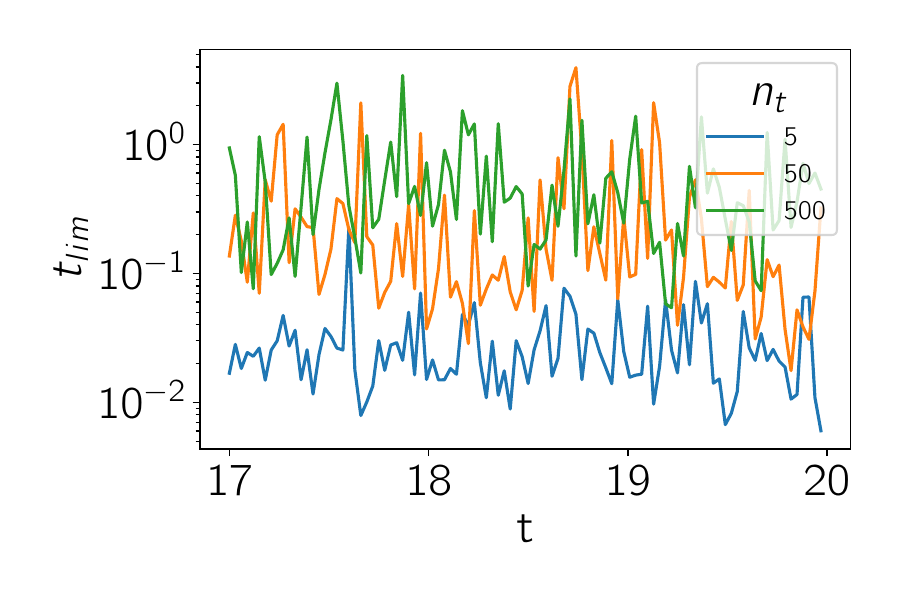}
  \caption{Limiting time-scale in the latent space ($t_{lim}$) (Eq.~\ref{eq: nODE tlim tmax}) as a function of time for neural ODEs with different training trajectory lengths $n_{t}$ ($N_w=100$ and four convolutional layers)}.
  \label{fig:2DAtmos_timeScaleComp}
\end{figure}

Figure~\ref{fig:2DAtmos_timeScaleComp} shows the evolution of the limiting time-scale in the latent space, $t_{lim}$, for neural ODEs trained with varying $n_t$ values. The results show that larger $n_t$ values correspond to increased $t_{lim}$, which promotes smoother trajectories in the latent space. This behavior is consistent with the trends observed for the KS equations in Figure~\ref{fig:tLim_nt} and the 1D channel detonation case in Figure~\ref{fig:1dDet_timeScaleComp}.

\subsubsection{Effect of time-step}

The effect of the time-step size \( \Delta t \), used in the latent space during inference, on the rollout error \( \mathcal{L}_{\text{RO}} \) is analyzed in this section. For this analysis, the best-performing model with \( n_t=20 \), trained using \( \Delta t = 10 \), was selected.

\begin{figure}
  \centering
  \includegraphics[width=0.49\textwidth]{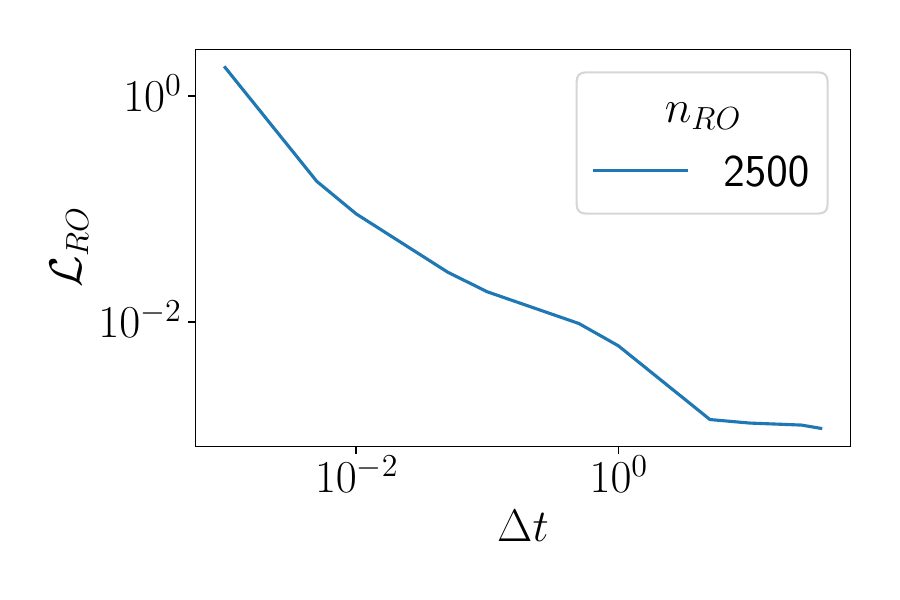}
  \caption{$\mathcal{L}_{\text{RO}}$ (Eq.~\ref{eq: RO_error}) for a rollout trajectory length $n_{\text{RO}} = 2500$, as a function of $\Delta_t$ ($N_w=100$ and four convolutional layers).}
  \label{fig:2DAtmos_LossVsDt}
\end{figure}

Figure~\ref{fig:2DAtmos_LossVsDt} illustrates the variation of the rollout error \( \mathcal{L}_{\text{RO}} \) as a function of the time-step \( \Delta t \) used in the latent space for a rollout trajectory length of \( n_{\text{RO}} = 1000 \). The results indicate that the error generally decreases as the time-step increases and eventually reaches a plateau. This behavior suggests that neural ODEs trained on discrete data points may primarily approximate discrete dynamics rather than capturing the underlying continuous dynamics. As a result, when the inference time-step deviates from the training time-step, the model's error can increase significantly. Notably, evaluating the model at different time-step sizes, whether larger or smaller than the training step, can lead to substantial errors, indicating potential overfitting to the training discretization and limited generalization across different time-step scales. Similar trends have also been observed in previous work~\cite{Krishnapriyan2023, mohan2024getneuralpartialdifferential, chakraborty2024noteerroranalysisdatadriven}, highlighting the challenges of generalizing across different time-step scales.

\color{black}

\section{Conclusions}

This work explores the applicability of autoencoder-based neural ODE surrogate models for accelerated simulation of PDE-based systems in which advection terms play a critical role. Overall, the strategy was found to produce effective models for challenging unsteady advection-dominated dynamics in the context of both the Kuramoto-Sivashinsky and compressible reacting Navier-Stokes equations (a detonation configuration described by detailed chemical kinetics). 

Alongside predictive demonstrations, however, the key thrust of this work was to uncover physical insight into the sources of model acceleration (i.e., how the neural ODE achieves its acceleration). Since acceleration in such surrogate models is intrinsically tied to the elimination of prohibitive time-scales, the conceptual framework to facilitate this portion of the study came from the quantification, and in-depth analysis, of neural ODE time-scales.

More specifically, through eigenvalue analysis of dynamical system Jacobians, the effects of both autoencoder and neural ODE components on latent system time-scales in the surrogate model was quantified and analyzed in this work. To this end, the sensitivity of various critical training parameters -- de-coupled versus end-to-end training, latent space dimensionality, and the role of training trajectory length, for example -- to both model accuracy and the discovered latent system timescales was investigated in detail. 

Conducting eigenvalue-based timescale analysis on the KS equations proved instrumental in isolating the impact of individual training hyperparameters on timescale elimination in the latent space. Notably, the number of convolutional layers in the autoencoder and the latent dimensionality exhibited little discernible effect on latent space timescales. Interestingly, the training methodology — in terms of coupled versus de-coupled optimization of the autoencoder and neural ODE components — also did not have any effect on the time-scale reduction, although, as reported in recent work \cite{Wan2023}, the decoupled training approach had lower single-step and rollout errors for both PDE systems tested here. 

On the other hand, the training trajectory length, denoted as \(n_t\), emerged as a crucial parameter governing latent timescales. More specifically, it was found that increasing \(n_t\) generally increases the limiting (smallest) timescales (\(t_{lim}\)) in latent space, resulting in smoother latent trajectories which in turn translates to greater levels of achievable acceleration. Exceeding a certain $n_t$ threshold, however, diminished predictive accuracy. Consequently, an optimal \(n_t\) value, striking a balance between timescale elimination and predictive accuracy, was identified for both the KS equation and the detonation test case.

Efforts to establish a framework for \textit{a priori} identification of this optimal \(n_t\) value based on the problem's physics involved analyzing the largest timescales (\(t_{max}\)) in the latent space for various \(n_t\) values. For the KS results observed here, the largest time-scales of the neural ODE exhibiting the best predictive accuracy were found to align closely with those of the full-order system, pointing to the role of capturing slow-moving dynamics in the surrogate latent space for enhanced accuracy. For the unsteady detonation {and atmospheric flow datasets}, a similar general trend was seen in that the rollout error was shown to be minimized at an particular value of $n_t$. Although promising, further studies need to be done to investigate both the generality of time-scale relationships observed here and neural ODE feasibility in modeling the compressible reacting Navier-Stokes equations {and atmospheric flows}. {Additionally, benchmarking different surrogate modeling architectures (such as DeepONets \cite{lu2021learning}, Fourier neural operators \cite{li2020fourier}, and other approaches), and assessing the transferability of the latent time scale trends uncovered in this work to these different architectures, is a useful direction.} Such aspects will be explored in future work.

\section*{Acknowledgments}
This research used resources of the Argonne Leadership Computing Facility (ALCF), a U.S. Department of Energy (DOE) Office of Science user facility at Argonne National Laboratory and is based on research supported by the U.S. DOE Office of Science-Advanced Scientific Computing Research Program, under Contract No. DE-AC02-06CH11357. The authors would also like to acknowledge the ALCF Summer intern program. SB and PP acknowledge laboratory-directed research and development (LDRD) funding support from Argonne’s Advanced Energy Technologies (AET) directorate through the Advanced Energy Technology and Security (AETS) Fellowship. RM acknowledges funding support from ASCR for DOE-FOA-2493 ``Data-intensive scientific machine learning’’.

% \printbibliography
\bibliography{refs}

\end{document}